\documentclass[aps,pre,reprint,tightenlines,twocolumn,showpacs,showkeys]{revtex4}
\usepackage[dvips]{graphicx}
\usepackage{hyperref,amsmath,amssymb,amsbsy}

\newcommand{\ii}{\textrm{i}}

\newcommand{\dd}{\textrm{d}}
\newcommand{\dupl}{\textrm{dupl.}}

\newcommand{\Tr}{\textrm{Tr}}
\newcommand{\bTr}{\textrm{bTr}}
\newcommand{\erfc}{\textrm{erfc}}
\newcommand{\Id}{\mathbf{1}}
\newcommand{\Zero}{\mathbf{0}}
\newcommand{\la}{\left\langle}
\newcommand{\ra}{\right\rangle}

\begin{document}

\title{Generalized Bures products from free probability}

\author{Andrzej \surname{Jarosz}}
\email{jedrekjarosz@gmail.com}
\affiliation{The Henryk Niewodnicza\'{n}ski Institute of Nuclear Physics, Polish Academy of Sciences, Radzikowskiego 152, 31-342 Krak\'{o}w, Poland}

\topmargin=0cm
\allowdisplaybreaks[4]

\begin{abstract}
Inspired by the theory of quantum information, I use two non-Hermitian random matrix models---a weighted sum of circular unitary ensembles and a product of rectangular Ginibre unitary ensembles---as building blocks of three new products of random matrices which are generalizations of the Bures model. I apply the tools of both Hermitian and non-Hermitian free probability to calculate the mean densities of their eigenvalues and singular values in the thermodynamic limit, along with their divergences at zero; the results are supported by Monte Carlo simulations. I pose and test conjectures concerning the relationship between the two densities (exploiting the notion of the $N$-transform), the shape of the mean domain of the eigenvalues (an extension of the single ring theorem), and the universal behavior of the mean spectral density close to the domain's borderline (using the complementary error function).
\end{abstract}


\maketitle


\section{Introduction}
\label{s:Introduction}

This paper is a continuation of the work~\cite{Jarosz2011-01}, whose focus was on the simplest characteristics (level density) of a weighted sum of unitary random matrices---a non-Hermitian random matrix model encountered e.g. in quantum information theory or the theory of random walks on trees. Inspired by the first of these areas of physics (Sec.~\ref{ss:Motivation}), I introduce more general models (Sec.~\ref{ss:Models}) and calculate---using the tools of free probability (Sec.~\ref{ss:Tools})---their level densities (Sec.~\ref{s:GeneralizedBuresProducts}).


\subsection{Models}
\label{ss:Models}


\subsubsection{Definitions of the models}
\label{sss:DefinitionsOfTheModels}

The goal of this article is a basic study of the following non-Hermitian random matrix models $\mathbf{T}$, $\mathbf{W}$, $\mathbf{V}$, which I will call ``generalized Bures products'' for the reason explained at the end of Sec.~\ref{sss:ApplicationsToQuantumInformationTheory}. In order to introduce them, I recall two other matrix models, $\mathbf{S}$ and $\mathbf{P}$, investigated recently in the literature:

(i) Let $L \geq 2$ be an integer; let \smash{$\mathbf{U}_{l}$}, for $l = 1 , 2 , \ldots , L$, be independent $N \times N$ unitary random matrices belonging to the Wigner-Dyson ``circular unitary ensemble'' (CUE or ``Haar measure''; i.e., their eigenvalues are on average uniformly distributed on the unit circle); let \smash{$w_{l}$} be arbitrary complex parameters. Then define a weighted sum,
\begin{equation}\label{eq:SDefinition}
\mathbf{S} \equiv w_{1} \mathbf{U}_{1} + w_{2} \mathbf{U}_{2} + \ldots + w_{L} \mathbf{U}_{L} .
\end{equation}
This non-Hermitian model may be referred to as the ``generalized Kesten model''~\cite{Kesten1959} (cf. also~\cite{HaagerupLarsen2000,GorlichJarosz2004}), and has been considered in some detail in~\cite{Jarosz2011-01}.

(ii) Let $K \geq 1$ be an integer; let \smash{$\mathbf{A}_{k}$}, for $k = 1 , 2 , \ldots , K$, be rectangular (of some dimensions \smash{$N_{k} \times N_{k + 1}$}) complex random matrices such that all the real and imaginary parts of the entries of each matrix are independent random numbers with the Gaussian distribution of zero mean and variance denoted by \smash{$\sigma_{k}^{2} / ( 2 ( N_{k} N_{k + 1} )^{1 / 2} )$}, i.e., the joint probability density function (JPDF),
\begin{equation}\label{eq:RectangularGinUEJPDF}
P \left( \{ \mathbf{A}_{k} \} \right) \propto \prod_{k = 1}^{K} \exp \left( - \frac{\sqrt{N_{k} N_{k + 1}}}{\sigma_{k}^{2}} \Tr \left( \mathbf{A}_{k}^{\dagger} \mathbf{A}_{k} \right) \right) ;
\end{equation}
these are rectangular ``Ginibre unitary ensembles'' (GinUE)~\cite{Ginibre1965,Girko19841985}. Then define their product, an \smash{$N_{1} \times N_{K + 1}$} random matrix,
\begin{equation}\label{eq:PDefinition}
\mathbf{P} \equiv \mathbf{A}_{1} \mathbf{A}_{2} \ldots \mathbf{A}_{K} .
\end{equation}
This model has been thoroughly investigated (cf.~e.g.~\cite{BanicaBelinschiCapitaineCollins2007,PensonZyczkowski2011,KanzieperSingh2010,BurdaJanikWaclaw2009,BurdaJaroszLivanNowakSwiech20102011}).

Now, having a number of copies of $\mathbf{S}$ and $\mathbf{P}$ (all of them assumed statistically independent from each other), one may use them as building blocks of various products:

(i) A product of $J \geq 1$ weighted sums (\ref{eq:SDefinition}),
\begin{equation}\label{eq:TDefinition}
\mathbf{T} \equiv \mathbf{S}_{1} \mathbf{S}_{2} \ldots \mathbf{S}_{J} ,
\end{equation}
where each sum \smash{$\mathbf{S}_{j}$} has arbitrary length \smash{$L_{j}$} and complex weights \smash{$w_{j l}$}.

(ii) A product of (\ref{eq:TDefinition}) and (\ref{eq:PDefinition}),
\begin{equation}\label{eq:WDefinition}
\mathbf{W} \equiv \mathbf{T} \mathbf{P} ,
\end{equation}
i.e., this is a string of $J$ generalized Kesten ensembles and $K$ Ginibre unitary ensembles. Note that one has to set \smash{$N = N_{1}$}, and then $\mathbf{W}$ is rectangular of dimensions \smash{$N_{1} \times N_{K + 1}$}.

(iii) Finally, a most general string of generalized Kesten and Ginibre unitary ensembles is obtained by multiplying a number $I \geq 1$ of random matrices (\ref{eq:WDefinition}),
\begin{equation}\label{eq:VDefinition}
\mathbf{V} \equiv \mathbf{W}_{1} \mathbf{W}_{2} \ldots \mathbf{W}_{I} .
\end{equation}
The dimensions of the terms (\smash{$N_{i , 1} \times N_{i , K_{i} + 1}$}, for $i = 1 , 2 , \ldots , I$) must obey \smash{$N_{i , K_{i} + 1} = N_{i + 1 , 1}$}, and then $\mathbf{V}$ is rectangular of dimensions \smash{$N_{1 , 1} \times N_{I , K_{I} + 1}$}.


\subsubsection{Thermodynamic limit}
\label{sss:ThermodynamicLimit}

The techniques I apply to approach the above models are valid only in the ``thermodynamic limit,'' i.e., for all the matrix dimensions infinite and their ``rectangularity ratios'' finite,
\begin{equation}
\begin{split}\label{eq:ThermodynamicLimit}
&N = N_{1} , N_{2} , \ldots , N_{K + 1} \to \infty ,\\
&r_{k} \equiv \frac{N_{k}}{N_{K + 1}} \textrm{ = finite.}
\end{split}
\end{equation}
[I use here the notation for the model $\mathbf{W}$ (\ref{eq:WDefinition}); for $\mathbf{V}$ (\ref{eq:VDefinition}), one should add an index $i$.]


\subsubsection{Mean densities of the eigenvalues and singular values}
\label{sss:MeanDensitiesOfTheEigenvaluesAndSingularValues}

\emph{Level densities.} This paper is just a most basic study of the models $\mathbf{X} \in \{ \mathbf{T} , \mathbf{W} , \mathbf{V} \}$, namely, I will be calculating only:

(i) The mean density of the eigenvalues (``mean spectral density'' or ``level density''),
\begin{equation}\label{eq:NonHermitianMeanSpectralDensityDefinition}
\rho_{\mathbf{X}} ( z , \overline{z} ) \equiv \frac{1}{N} \sum_{i = 1}^{N} \la \delta^{( 2 )} \left( z - \lambda_{i} \right) \ra ,
\end{equation}
where the mean values are with respect to the JPDF of $\mathbf{X}$ and are evaluated at the complex Dirac delta functions at the (generically complex) eigenvalues \smash{$\lambda_{i}$} of $\mathbf{X}$. Note that one must set \smash{$N_{K + 1} = N$} (i.e., \smash{$r_{1} = 1$}) for $\mathbf{W}$ to be a square matrix.

(ii) The mean density of the ``singular values,'' which are the (real and non-negative) eigenvalues \smash{$\mu_{i}$} of the \smash{$N_{K + 1} \times N_{K + 1}$} Hermitian random matrix \smash{$\mathbf{H} \equiv \mathbf{X}^{\dagger} \mathbf{X}$},
\begin{equation}\label{eq:HermitianMeanSpectralDensityDefinition}
\rho_{\mathbf{H}} ( x ) \equiv \frac{1}{N_{K + 1}} \sum_{i = 1}^{N_{K + 1}} \la \delta \left( x - \mu_{i} \right) \ra ,
\end{equation}
where the Dirac delta is real. This time, \smash{$r_{1}$} may acquire any positive value.

\emph{Future work---universal quantities.} Certainly, this is but the first step in understanding the considered models. The level density of a random matrix is known not to be universal, i.e., it depends on the precise probability distribution of the matrix. (However, cf.~the end of Sec.~\ref{sss:FreeProbabilityAndModelP} for a hint in favor of a certain universality of the level densities of our models.) As a next step, it would be desirable to investigate some ``universal'' properties of our models, i.e., depending solely on their symmetries but not the specific probability distributions. Some basic universal observables would be a ``two-point connected correlation function,''
\begin{subequations}
\begin{align}
\rho^{\textrm{connected}}_{\mathbf{H}} ( x , y ) &\equiv \rho_{\mathbf{H}} ( x , y ) - \rho_{\mathbf{H}} ( x ) \rho_{\mathbf{H}} ( y ) ,\label{eq:HermitianTwoPointConnectedCorrelationFunctionDefinition1}\\
\rho_{\mathbf{H}} ( x , y ) &\equiv \frac{1}{N_{K + 1}^{2}} \sum_{i , j = 1}^{N_{K + 1}} \la \delta \left( x - \mu_{i} \right) \delta \left( y - \mu_{j} \right) \ra\label{eq:HermitianTwoPointConnectedCorrelationFunctionDefinition2}
\end{align}
\end{subequations}
(here written in the Hermitian case), or the behavior of the level density/correlation function close to the borderline (for non-Hermitian models) or the edges (for Hermitian models) of the spectrum.

\emph{Universal erfc scaling.} Even though I focus on the level densities in the bulk of the spectrum and not close to its borders, the models $\mathbf{T}$, $\mathbf{W}$, $\mathbf{V}$ share a certain property (their mean spectrum is rotationally-symmetric around zero; cf.~Sec.~\ref{sss:RotationalSymmetryAndNTransformConjecture}) which permits an application of the so-called ``erfc conjecture'' (cf.~Sec.~\ref{sss:ErfcConjecture}), which describes the universal way in which the mean spectral density (\ref{eq:NonHermitianMeanSpectralDensityDefinition}) is modified close to the borderline of the ``mean spectral domain'' $\mathcal{D}$ (i.e., the subset of the complex plane where the eigenvalues of an infinite random matrix fall). The pertinent form-factor (\ref{eq:FiniteSizeBorderlineFactor}) depends on one or two parameters, and I positively test this hypothesis on a number of examples for each class of the models $\mathbf{T}$, $\mathbf{W}$, $\mathbf{V}$ by fitting these parameters to match the Monte Carlo data. However, it still remains to prove this hypothesis, as well as analytically determine the form-factor parameters.

\emph{Future work---microscopic quantities.} Another step in the research on our models could be to consider finite matrix dimensions, and to calculate the complete JPDF of the eigenvalues and singular values of $\mathbf{S}$, $\mathbf{P}$, $\mathbf{T}$, $\mathbf{W}$, $\mathbf{V}$. This however is a much more involved task.


\subsection{Tools}
\label{ss:Tools}

In this Section, I sketch the means by which the level densities [(\ref{eq:NonHermitianMeanSpectralDensityDefinition}), (\ref{eq:HermitianMeanSpectralDensityDefinition})] of the models [(\ref{eq:TDefinition}), (\ref{eq:WDefinition}), (\ref{eq:VDefinition})] will be evaluated in the thermodynamic limit (\ref{eq:ThermodynamicLimit}). I do not delve into details, as they have been addressed e.g.~in~\cite{Jarosz2011-01,Jarosz2010-01}.


\subsubsection{Basic language of random matrix theory}
\label{sss:BasicLanguageOfRandomMatrixTheory}

\emph{Mean spectral densities and Green functions.} First of all, it is convenient to replace the mean spectral density of any ($N \times N$, $N \to \infty$) Hermitian random matrix $\mathbf{H}$ or non-Hermitian random matrix $\mathbf{X}$ with an equivalent but more tractable object in the following way: Since the definitions [(\ref{eq:HermitianMeanSpectralDensityDefinition}), (\ref{eq:NonHermitianMeanSpectralDensityDefinition})] exploit the real (Hermitian case) or complex (non-Hermitian case) Dirac delta function, one uses their respective representations,
\begin{subequations}
\begin{align}
\delta ( x ) &= - \frac{1}{2 \pi \ii} \lim_{\epsilon \to 0^{+}} \left( \frac{1}{x + \ii \epsilon} - \frac{1}{x - \ii \epsilon} \right) ,\label{eq:RealDiracDeltaDefinition}\\
\delta^{( 2 )} ( z ) &= \frac{1}{\pi} \partial_{\overline{z}} \lim_{\epsilon \to 0} \frac{\overline{z}}{| z |^{2} + \epsilon^{2}} ,\label{eq:ComplexDiracDeltaDefinition}
\end{align}
\end{subequations}
which prompt one to introduce the following ``holomorphic'' or ``nonholomorphic Green functions (resolvents)''~\cite{SommersCrisantiSompolinskyStein1988,HaakeIzrailevLehmannSaherSommers1992,LehmannSaherSokolovSommers1995,FyodorovSommers1997,FyodorovKhoruzhenkoSommers1997},
\begin{subequations}
\begin{align}
G_{\mathbf{H}} ( z ) &\equiv \frac{1}{N} \sum_{i = 1}^{N} \la \frac{1}{z - \mu_{i}} \ra = \nonumber\\
&= \frac{1}{N} \Tr \la \frac{1}{z \Id_{N} - \mathbf{H}} \ra ,\label{eq:HolomorphicGreenFunctionDefinition}\\
G_{\mathbf{X}} ( z , \overline{z} ) &\equiv \lim_{\epsilon \to 0} \lim_{N \to \infty} \frac{1}{N} \sum_{i = 1}^{N} \la \frac{\overline{z} - \overline{\lambda_{i}}}{\left| z - \lambda_{i} \right|^{2} + \epsilon^{2}} \ra =\nonumber\\
&= \lim_{\epsilon \to 0} \lim_{N \to \infty} \frac{1}{N} \Tr \cdot \nonumber\\
&\cdot \la \frac{\overline{z} \Id_{N} - \mathbf{X}^{\dagger}}{\left( z \Id_{N} - \mathbf{X} \right) \left( \overline{z} \Id_{N} - \mathbf{X}^{\dagger} \right) + \epsilon^{2} \Id_{N}} \ra .\label{eq:NonHolomorphicGreenFunctionDefinition}
\end{align}
\end{subequations}
Thanks to (\ref{eq:RealDiracDeltaDefinition})-(\ref{eq:ComplexDiracDeltaDefinition}), the densities are straightforward to reproduce from them,
\begin{subequations}
\begin{align}
\rho_{\mathbf{H}} ( x ) &= - \frac{1}{2 \pi \ii} \lim_{\epsilon \to 0^{+}} \left( G_{\mathbf{H}} ( x + \ii \epsilon ) - G_{\mathbf{H}} ( x - \ii \epsilon ) \right) ,\label{eq:MeanSpectralDensityFromHolomorphicGreenFunction}\\
\rho_{\mathbf{X}} ( z , \overline{z} ) &= \frac{1}{\pi} \partial_{\overline{z}} G_{\mathbf{X}} ( z , \overline{z} ) ,\label{eq:MeanSpectralDensityFromNonHolomorphicGreenFunction}
\end{align}
\end{subequations}
while the resolvents prove to be handier from the computational point of view.

Remark: Eqs.~[(\ref{eq:NonHolomorphicGreenFunctionDefinition}), (\ref{eq:MeanSpectralDensityFromNonHolomorphicGreenFunction})] are valid for $z$ inside the mean spectral domain $\mathcal{D}$. Outside it, the regulator $\epsilon$ may be safely set to zero, and the nonholomorphic Green function reduces to its holomorphic counterpart (\ref{eq:HolomorphicGreenFunctionDefinition}),
\begin{equation}\label{eq:NonHolomorphicGreenFunctionOutsideD}
G_{\mathbf{X}} ( z , \overline{z} ) = G_{\mathbf{X}} ( z ) , \quad \textrm{for } z \notin \mathcal{D} .
\end{equation}
An implication is that by calculating the Green function both inside $\mathcal{D}$ and outside $\mathcal{D}$, and then equating them according to (\ref{eq:NonHolomorphicGreenFunctionOutsideD}), one arrives at an equation of the borderline $\partial \mathcal{D}$.

\emph{$M$-transforms.} In addition to the resolvents, one often finds even more convenient to work with their simple modification, the ``holomorphic'' or ``nonholomorphic $M$-transforms,''
\begin{subequations}
\begin{align}
M_{\mathbf{H}} ( z ) &\equiv z G_{\mathbf{H}} ( z ) - 1 ,\label{eq:HolomorphicMTransformDefinition}\\
M_{\mathbf{X}} ( z , \overline{z} ) &\equiv z G_{\mathbf{X}} ( z , \overline{z} ) - 1 .\label{eq:NonHolomorphicMTransformDefinition}
\end{align}
\end{subequations}
The Hermitian $M$-transform (\ref{eq:HolomorphicMTransformDefinition}) has the interpretation of the generating function of the ``moments'' \smash{$m_{\mathbf{H} , n}$} (if they exist), by expanding around $z = \infty$,
\begin{equation}\label{eq:HolomorphicMTransformMomentExpansion}
M_{\mathbf{H}} ( z ) = \sum_{n \geq 1} \frac{m_{\mathbf{H} , n}}{z^{n}} , \quad m_{\mathbf{H} , n} \equiv \frac{1}{N} \Tr \la \mathbf{H}^{n} \ra .
\end{equation}


\subsubsection{Rotational symmetry and $N$-transform conjecture}
\label{sss:RotationalSymmetryAndNTransformConjecture}

\emph{Rotational symmetry of the mean spectrum.} In this paper, I investigate only non-Hermitian random matrix models $\mathbf{X}$ with a feature that their mean spectrum is rotationally-symmetric around zero, i.e., that their density (\ref{eq:NonHermitianMeanSpectralDensityDefinition}) depends only on
\begin{equation}\label{eq:RDefinition}
R \equiv | z | .
\end{equation}
Equivalently, this means that their nonholomorphic $M$-transform (\ref{eq:NonHolomorphicMTransformDefinition}) also depends only on the radius,
\begin{equation}\label{eq:RotationallySymmetricNonHolomorphicMTransformDefinition}
M_{\mathbf{X}} ( z , \overline{z} ) = \mathfrak{M}_{\mathbf{X}} \left( R^{2} \right) ,
\end{equation}
giving the relevant radial part of the density through
\begin{equation}\label{eq:RadialMeanSpectralDensityFromRotationallySymmetricNonHolomorphicMTransform}
\rho^{\textrm{rad.}}_{\mathbf{X}} ( R ) \equiv 2 \pi R \left. \rho_{\mathbf{X}} ( z , \overline{z} ) \right|_{| z | = R} = \frac{\dd}{\dd R} \mathfrak{M}_{\mathbf{X}} \left( R^{2} \right) .
\end{equation}

\emph{$N$-transform conjecture.} For such ensembles, it has been suggested and numerically confirmed on a number of examples~\cite{BurdaJaroszLivanNowakSwiech20102011,Jarosz2011-01,Jarosz2010-01} that there exists a simple relationship between the mean densities of their eigenvalues and singular values (i.e., eigenvalues of \smash{$\mathbf{X}^{\dagger} \mathbf{X}$})---the ``$N$-transform conjecture.'' In order to express it, one needs to define the functional inverse of the $M$-transform. In the Hermitian case (\ref{eq:HolomorphicMTransformDefinition}), it can be directly done,
\begin{equation}\label{eq:HolomorphicNTransformDefinition}
M_{\mathbf{H}} \left( N_{\mathbf{H}} ( z ) \right) = N_{\mathbf{H}} \left( M_{\mathbf{H}} ( z ) \right) = z ,
\end{equation}
obtaining the ``holomorphic $N$-transform.'' In a generic non-Hermitian case (\ref{eq:NonHolomorphicMTransformDefinition}), however, such an inversion is obviously impossible. But with the rotational symmetry present (\ref{eq:RotationallySymmetricNonHolomorphicMTransformDefinition}), the $M$-transform again depends on one argument, and the inversion becomes generically doable,
\begin{equation}\label{eq:RotationallySymmetricNonHolomorphicNTransformDefinition}
\mathfrak{M}_{\mathbf{X}} \left( \mathfrak{N}_{\mathbf{X}} ( z ) \right) = z , \quad \mathfrak{N}_{\mathbf{X}} \left( \mathfrak{M}_{\mathbf{X}} \left( R^{2} \right) \right) = R^{2} ,
\end{equation}
which is called the ``rotationally-symmetric nonholomorphic $N$-transform.'' [To be more precise, \smash{$\mathfrak{N}_{\mathbf{X}} ( z )$} defined by the left equation in (\ref{eq:RotationallySymmetricNonHolomorphicNTransformDefinition}) is a holomorphic continuation of the one defined by the right equation.] Now the hypothesis states that these $N$-transforms of $\mathbf{X}$ and \smash{$\mathbf{H} \equiv \mathbf{X}^{\dagger} \mathbf{X}$} remain in a simple relation,
\begin{equation}\label{eq:NTransformConjecture}
N_{\mathbf{X}^{\dagger} \mathbf{X}} ( z ) = \frac{z + 1}{z} \mathfrak{N}_{\mathbf{X}} ( z ) .
\end{equation}

Typically, one of these random matrices will be much more accessible by analytical methods than the other, and therefore, Eq.~(\ref{eq:NTransformConjecture}) will provide a way to that more complicated model.

I will henceforth make extensive use of this hypothesis; numerical tests of the results thus obtained will further support it.


\subsubsection{Free probability and model $\mathbf{P}$}
\label{sss:FreeProbabilityAndModelP}

\emph{Algorithm for computing the densities of $\mathbf{P}$ and \smash{$\mathbf{P}^{\dagger} \mathbf{P}$}.} The model $\mathbf{P}$ (\ref{eq:PDefinition}) is a product of independent (rectangular) random matrices \smash{$\mathbf{A}_{k}$}. If \smash{$r_{1} \neq 1$} (i.e., $\mathbf{P}$ is rectangular), then one is interested in the eigenvalues of \smash{$\mathbf{P}^{\dagger} \mathbf{P}$} only, while if \smash{$r_{1} = 1$} (i.e., $\mathbf{P}$ is square), then its mean spectrum has rotational symmetry around zero (\ref{eq:RotationallySymmetricNonHolomorphicMTransformDefinition})~\cite{BurdaJanikWaclaw2009,BurdaJaroszLivanNowakSwiech20102011}, and can therefore be related (\ref{eq:NTransformConjecture}) to the spectrum of \smash{$\mathbf{P}^{\dagger} \mathbf{P}$}. The level density of this latter matrix can now be expressed (by cyclic shifts of the terms)~\cite{BurdaJaroszLivanNowakSwiech20102011} through the density of the product of the Hermitian (``Wishart''~\cite{Wishart1928}) matrices \smash{$\mathbf{A}_{k}^{\dagger} \mathbf{A}_{k}$}. Hence, the situation is suitable for the employment of the ``multiplication law'' of free probability calculus of Voiculescu and Speicher~\cite{VoiculescuDykemaNica1992,Speicher1994}.

\emph{Multiplication law in free probability.} This theory is essentially a probability theory of noncommuting objects, such as large random matrices. Its foundational notion is that of ``freeness,'' being a proper generalization of statistical independence. Qualitatively speaking, freeness requires the matrices \smash{$\mathbf{H}_{1}$}, \smash{$\mathbf{H}_{2}$} to not only be independent statistically, but also ``rotationally'' (i.e., no distinguished direction in their probability measures, i.e., dependence only on the eigenvalues)---independent random rotations from the CUE, \smash{$\mathbf{U}_{1} \mathbf{H}_{1} \mathbf{U}_{1}^{\dagger}$} and \smash{$\mathbf{U}_{2} \mathbf{H}_{2} \mathbf{U}_{2}^{\dagger}$}, would ensure that condition if necessary.

Assuming freeness, there is a very useful prescription for computing the mean spectral density of the product \smash{$\mathbf{H}_{1} \mathbf{H}_{2}$} (assuming it is still Hermitian, which obviously is not always the case; but the same procedure applies when all the matrices are unitary), once the densities of the terms are known: (i) Encode the densities by the holomorphic $M$-transforms (\ref{eq:HolomorphicMTransformDefinition}), \smash{$M_{\mathbf{H}_{1}} ( z )$} and \smash{$M_{\mathbf{H}_{2}} ( z )$}, as outlined in Sec.~\ref{sss:BasicLanguageOfRandomMatrixTheory}. (ii) Invert them functionally to find the respective $N$-transforms (\ref{eq:HolomorphicNTransformDefinition}), \smash{$N_{\mathbf{H}_{1}} ( z )$} and \smash{$N_{\mathbf{H}_{2}} ( z )$}. (iii) The $N$-transform of the product follows from the ``multiplication law,''
\begin{equation}\label{eq:MultiplicationLaw}
N_{\mathbf{H}_{1} \mathbf{H}_{2}} ( z ) = \frac{z}{z + 1} N_{\mathbf{H}_{1}} ( z ) N_{\mathbf{H}_{2}} ( z ) .
\end{equation}
(iv) Invert the result functionally to find the $M$-transform, and thus the mean spectral density of the product.

\emph{Eigenvalues and singular values of $\mathbf{P}$.} Implementing the above steps to $\mathbf{P}$, one readily discovers~\cite{BurdaJaroszLivanNowakSwiech20102011},
\begin{equation}\label{eq:HolomorphicNTransformOfPDaggerP}
N_{\mathbf{P}^{\dagger} \mathbf{P}} ( z ) = \sigma^{2} \sqrt{r_{1}} \frac{z + 1}{z} \prod_{k = 1}^{K} \left( \frac{z}{r_{k}} + 1 \right) ,
\end{equation}
where for short, \smash{$\sigma \equiv \prod_{k = 1}^{K} \sigma_{k}$}, from which stems both the mean density of the singular values [(\ref{eq:HolomorphicNTransformDefinition}), (\ref{eq:HolomorphicMTransformDefinition}), (\ref{eq:MeanSpectralDensityFromHolomorphicGreenFunction})] and (provided that \smash{$r_{1} = 1$}) the eigenvalues [(\ref{eq:NTransformConjecture}), (\ref{eq:RotationallySymmetricNonHolomorphicNTransformDefinition}), (\ref{eq:RadialMeanSpectralDensityFromRotationallySymmetricNonHolomorphicMTransform})] of $\mathbf{P}$ [in the latter case, the mean spectral domain $\mathcal{D}$ is a centered disk of radius \smash{$R_{\textrm{ext.}} = \sigma$}, cf.~(\ref{eq:XRExt})-(\ref{eq:XRInt})].

Furthermore, the authors of~\cite{BurdaJanikWaclaw2009} conjecture and numerically affirm that (\ref{eq:HolomorphicNTransformOfPDaggerP}) is to some degree universal, namely that it holds for the matrix elements of the \smash{$\mathbf{A}_{k}$} not only IID Gaussian but also arbitrary IID obeying the Pastur-Lindeberg condition. Depending on a possible analogous universality of the level density of $\mathbf{S}$, also the densities of $\mathbf{T}$, $\mathbf{W}$, $\mathbf{V}$ may exhibit a certain universality.

Let me close by saying that an algorithm parallel to the one described above will also be applied to derive the main results of this paper (cf.~Sec.~\ref{ss:GeneralizedMultiplicationLaws}).


\subsubsection{Quaternion free probability and model $\mathbf{S}$}
\label{sss:QuaternionFreeProbabilityAndModelS}

Although the topic of this work is multiplication rather than summation of random matrices, I should mention another [in addition to the technique outlined in Sec.~\ref{sss:FreeProbabilityAndModelP} and Eq.~(\ref{eq:HolomorphicNTransformOfPDaggerP})] pillar of this article---the results for the model $\mathbf{S}$ (\ref{eq:SDefinition}) derived in~\cite{Jarosz2011-01}.

\emph{Addition law in free probability.} In the realm of Hermitian random matrices, free probability provides---besides the multiplication law (\ref{eq:MultiplicationLaw})---a rule for summing free matrices \smash{$\mathbf{H}_{1}$}, \smash{$\mathbf{H}_{2}$}. This time, one should invert functionally not the holomorphic $M$-transforms, but the Green functions (\ref{eq:HolomorphicGreenFunctionDefinition}),
\begin{equation}\label{eq:HolomorphicBlueFunctionDefinition}
G_{\mathbf{H}} \left( B_{\mathbf{H}} ( z ) \right) = B_{\mathbf{H}} \left( G_{\mathbf{H}} ( z ) \right) = z ,
\end{equation}
which is known as the ``holomorphic Blue function''~\cite{Zee1996}. This objects then satisfies the ``addition law''~\cite{VoiculescuDykemaNica1992,Speicher1994},
\begin{equation}\label{eq:HermitianAdditionLaw}
B_{\mathbf{H}_{1} + \mathbf{H}_{2}} ( z ) = B_{\mathbf{H}_{1}} ( z ) + B_{\mathbf{H}_{2}} ( z ) - \frac{1}{z} ,
\end{equation}
upon which it remains to functionally invert the left-hand side, which leads to the holomorphic Green function of the sum, and consequently, its mean spectral density. (Recall that in classical probability theory, an analogous algorithm makes use of the logarithms of the characteristic functions of independent random variables.)

\emph{Addition law in quaternion free probability.} For non-Hermitian random matrices, a construction parallel to (\ref{eq:HolomorphicBlueFunctionDefinition})-(\ref{eq:HermitianAdditionLaw}) has been worked out in~\cite{JaroszNowak2004,JaroszNowak2006}. Basically, it consists of three steps:

(i) ``Duplication''~\cite{JanikNowakPappWambachZahed1997,JanikNowakPappZahed1997-01} (cf.~also~\cite{FeinbergZee1997-01,FeinbergZee1997-02})---the nonholomorphic Green function (\ref{eq:NonHolomorphicGreenFunctionDefinition}) has a denominator quadratic in $\mathbf{X}$, which makes its evaluation hard, and one needs to linearize it by introducing the ($2 \times 2$) ``matrix-valued Green function,''
\begin{equation}\label{eq:MatrixValuedGreenFunctionDefinition1}
\mathcal{G}_{\mathbf{X}} ( z , \overline{z} ) \equiv \lim_{\epsilon \to 0} \lim_{N \to \infty} \frac{1}{N} \bTr \la \frac{1}{\mathcal{Z}_{\epsilon} \otimes \Id_{N} - \mathbf{X}^{\dupl}} \ra ,
\end{equation}
where for short,
\begin{equation}\label{eq:MatrixValuedGreenFunctionDefinition2}
\mathcal{Z}_{\epsilon} \equiv \left( \begin{array}{cc} z & \ii \epsilon \\ \ii \epsilon & \overline{z} \end{array} \right) , \quad \mathbf{X}^{\dupl} \equiv \left( \begin{array}{cc} \mathbf{X} & \Zero_{N} \\ \Zero_{N} & \mathbf{X}^{\dagger} \end{array} \right) ,
\end{equation}
and the ``block--trace,''
\begin{equation}\label{eq:BlockTraceDefinition}
\bTr \left( \begin{array}{cc} \mathbf{A} & \mathbf{B} \\ \mathbf{C} & \mathbf{D} \end{array} \right) \equiv \left( \begin{array}{cc} \Tr \mathbf{A} & \Tr \mathbf{B} \\ \Tr \mathbf{C} & \Tr \mathbf{D} \end{array} \right) .
\end{equation}
In other words, this object resembles the holomorphic Green function (\ref{eq:HolomorphicGreenFunctionDefinition}), and thus can be approached by methods designed for Hermitian random matrices; the cost being the need to work with $2 \times 2$ matrices rather than complex numbers. The nonholomorphic Green function lies precisely on its upper left entry, \smash{$[ \mathcal{G}_{\mathbf{X}} ( z , \overline{z} ) ]_{1 1} = G_{\mathbf{X}} ( z , \overline{z} )$}.

(ii) The extension to the complete quaternion space---by replacing in (\ref{eq:MatrixValuedGreenFunctionDefinition1}) the infinitesimal $\epsilon$ with a finite complex number,
\begin{equation}\label{eq:QuaternionDefinition}
\mathcal{Q} \equiv \left( \begin{array}{cc} c & \ii \overline{d} \\ \ii d & \overline{c} \end{array} \right) , \quad c , d \in \mathbb{C} ,
\end{equation}
one obtains a quaternion argument of the ``quaternion Green function,''
\begin{equation}\label{eq:QuaternionGreenFunctionDefinition}
\mathcal{G}_{\mathbf{X}} ( \mathcal{Q} ) \equiv \frac{1}{N} \bTr \la \frac{1}{\mathcal{Q} \otimes \Id_{N} - \mathbf{X}^{\dupl}} \ra .
\end{equation}
The mean spectral density is reproduced from this quaternion function by approaching the complex plane ($c = z$, $d = \epsilon$), just as in the Hermitian case, it is obtained from the complex Green function by approaching the real line ($z = x \pm \ii \epsilon$) (\ref{eq:MeanSpectralDensityFromHolomorphicGreenFunction}).

(iii) The functional inversion can now be performed in the quaternion space,
\begin{equation}\label{eq:QuaternionBlueFunctionDefinition}
\mathcal{G}_{\mathbf{X}} \left( \mathcal{B}_{\mathbf{X}} ( \mathcal{Q} ) \right) = \mathcal{B}_{\mathbf{X}} \left( \mathcal{G}_{\mathbf{X}} ( \mathcal{Q} ) \right) = \mathcal{Q}
\end{equation}
(the ``quaternion Blue function''), and the ``quaternion addition law'' for free non-Hermitian random matrices can be proven,
\begin{equation}\label{eq:QuaternionAdditionLaw}
\mathcal{B}_{\mathbf{X}_{1} + \mathbf{X}_{2}} ( \mathcal{Q} ) = \mathcal{B}_{\mathbf{X}_{1}} ( \mathcal{Q} ) + \mathcal{B}_{\mathbf{X}_{2}} ( \mathcal{Q} ) - \mathcal{Q}^{- 1} .
\end{equation}

\emph{Eigenvalues and singular values of $\mathbf{S}$.} Formula~(\ref{eq:QuaternionAdditionLaw}) massively simplifies calculations of the mean spectral density of sums of free non-Hermitian ensembles. In particular, harnessing it to the weighted sum (\ref{eq:SDefinition}) implies that: (i) Its mean spectrum exhibits the rotational symmetry around zero (\ref{eq:RotationallySymmetricNonHolomorphicMTransformDefinition}). (ii) Its rotationally-symmetric nonholomorphic $N$-transform (\ref{eq:RotationallySymmetricNonHolomorphicNTransformDefinition}) is a solution to the following set of $( L + 2 )$ polynomial equations,
\begin{subequations}
\begin{align}
z &= \sum_{l = 1}^{L} M_{l} ,\label{eq:SMasterEquation1}\\
- C &= \frac{z ( z + 1 )}{\mathfrak{N}_{\mathbf{S}} ( z )} ,\label{eq:SMasterEquation2}\\
- C &= \frac{M_{l} \left( M_{l} + 1 \right)}{\left| w_{l} \right|^{2}} , \quad l = 1 , 2 , \ldots , L ,\label{eq:SMasterEquation3}
\end{align}
\end{subequations}
where $C \geq 0$ and \smash{$M_{l}$} are auxiliary unknowns. (iii) Its mean spectral domain $\mathcal{D}$ is either a disk or an annulus, whose external and internal radii are found from
\begin{subequations}
\begin{align}
R_{\textrm{ext.}}^{2} &= \mathfrak{N}_{\mathbf{S}} ( 0 ) ,\label{eq:SRExt}\\
R_{\textrm{int.}}^{2} &= \mathfrak{N}_{\mathbf{S}} ( - 1 ) .\label{eq:SRInt}
\end{align}
\end{subequations}
Furthermore, Eq.~(\ref{eq:NTransformConjecture}) means that the singular values are derived from an identical set of equations except (\ref{eq:SMasterEquation2}) which turns into
\begin{equation}\label{eq:SdSMasterEquation2}
- C = \frac{( z + 1 )^{2}}{N_{\mathbf{S}^{\dagger} \mathbf{S}} ( z )} .
\end{equation}


\subsubsection{Single ring conjecture}
\label{sss:SingleRingConjecture}

\emph{Single ring conjecture.} It is another hypothesis~\cite{Jarosz2011-01} (being a generalization of the ``Feinberg-Zee single ring theorem''~\cite{FeinbergZee1997-02,FeinbergScalettarZee2001,Feinberg2006,GuionnetKrishnapurZeitouni2009}) that for non-Hermitian random matrix models with the rotational symmetry (\ref{eq:RotationallySymmetricNonHolomorphicMTransformDefinition}) present, their mean spectral domain $\mathcal{D}$ is always a disk or an annulus. It has been proven for the models $\mathbf{S}$ and $\mathbf{P}$; for $\mathbf{T}$, $\mathbf{W}$, $\mathbf{V}$, I will assume it holds, and support it a posteriori by numerical simulations.

\emph{Radii of the disk/annulus.} If this conjecture is true, then the external and internal radii of the annulus (becoming a disk if the internal radius shrinks to zero) are still given by (\ref{eq:SRExt})-(\ref{eq:SRInt}), provided that there are no zero modes. Indeed, on the borderline of the domain $\mathcal{D}$, the nonholomorphic Green function (from the inside of $\mathcal{D}$) and the holomorphic one (from the outside of $\mathcal{D}$) must match (\ref{eq:NonHolomorphicGreenFunctionOutsideD}). But the holomorphic Green function compatible with the rotational symmetry (\ref{eq:RotationallySymmetricNonHolomorphicMTransformDefinition}) has the form \smash{$G_{\mathbf{X}} ( z ) = ( 1 + \mathfrak{M}_{\mathbf{X}} ( R^{2} ) ) / z$}, hence, holomorphicity leaves one possibility, i.e., that \smash{$\mathfrak{M}_{\mathbf{X}} ( R^{2} )$} is a constant, \smash{$G_{\mathbf{X}} ( z ) = ( 1 + \mathfrak{M} ) / z$}. In the external outside of $\mathcal{D}$ (including $z = \infty$), it is well-known that \smash{$G_{\mathbf{X}} ( z ) \sim 1 / z$} as $z \to \infty$, which means that $\mathfrak{M} = 0$. In the internal outside (including $z = 0$), supposing there are no zero modes, the Green function must be just zero, i.e., $\mathfrak{M} = - 1$. If there are zero modes, \smash{$\rho^{\textrm{zero modes}}_{\mathbf{X}} ( z , \overline{z} ) = \alpha \delta^{( 2 )} ( z , \overline{z} )$}, they are obtained by applying (\ref{eq:MeanSpectralDensityFromNonHolomorphicGreenFunction}) to the Green function \smash{$G^{\textrm{zero modes}}_{\mathbf{X}} ( z ) = \alpha / z$} regularized according to Eq.~(\ref{eq:ComplexDiracDeltaDefinition}); therefore, $\mathfrak{M} = \alpha - 1$. In other words,
\begin{subequations}
\begin{align}
R_{\textrm{ext.}}^{2} &= \mathfrak{N}_{\mathbf{X}} ( 0 ) ,\label{eq:XRExt}\\
R_{\textrm{int.}}^{2} &= \mathfrak{N}_{\mathbf{X}} ( \alpha - 1 ) .\label{eq:XRInt}
\end{align}
\end{subequations}


\subsubsection{erfc conjecture}
\label{sss:ErfcConjecture}

Calculations in the thermodynamic limit (\ref{eq:ThermodynamicLimit}) are capable of reproducing the mean spectral density only in the bulk of the domain $\mathcal{D}$, but not close to its borderline. However, based on earlier works~\cite{ForresterHonner1999,Kanzieper2005,KhoruzhenkoSommers2009,KanzieperSingh2010}, it has been suggested~\cite{BurdaJaroszLivanNowakSwiech20102011,Jarosz2011-01,Jarosz2010-01} how to extend the radial mean spectral density (\ref{eq:RadialMeanSpectralDensityFromRotationallySymmetricNonHolomorphicMTransform}) to the vicinity of the borderline, provided that the rotational symmetry (\ref{eq:RotationallySymmetricNonHolomorphicMTransformDefinition}) holds.

For each circle \smash{$R = R_{\textrm{b}}$} enclosing $\mathcal{D}$ (one or two; cf.~Sec.~\ref{sss:SingleRingConjecture}), one should simply multiply the radial density \smash{$\rho^{\textrm{rad.}}_{\mathbf{X}} ( R )$} by the universal form-factor,
\begin{equation}\label{eq:FiniteSizeBorderlineFactor}
f_{N , q_{\textrm{b}} , R_{\textrm{b}} , s_{\textrm{b}}} ( R ) \equiv \frac{1}{2} \erfc \left( q_{\textrm{b}} s_{\textrm{b}} \left( R - R_{\textrm{b}} \right) \sqrt{N} \right) ,
\end{equation}
where \smash{$\erfc ( x ) \equiv \frac{2}{\sqrt{\pi}} \int_{x}^{\infty} \dd t \exp ( - t^{2} )$} is the complementary error function, while the sign \smash{$s_{\textrm{b}}$} is $+ 1$ for the external borderline and $- 1$ for the internal borderline, and \smash{$q_{\textrm{b}}$} is a parameter dependent on the particular model, whose evaluation requires truly finite-size methods, but whose value I will adjust by fitting to the Monte Carlo data.


\subsection{Motivation}
\label{ss:Motivation}


\subsubsection{Interesting mathematical properties}
\label{sss:InterestingMathematicalProperties}

I find the models $\mathbf{T}$, $\mathbf{W}$, $\mathbf{V}$ mathematically interesting for the following reasons: (i) They are non-Hermitian, and the theory of such random matrices is richer and much less developed than for the Hermitian ones (cf.~e.g.~\cite{KhoruzhenkoSommers2009}). (ii) They belong to a special class of non-Hermitian matrices, namely, with rotationally-symmetric mean spectrum (\ref{eq:RotationallySymmetricNonHolomorphicMTransformDefinition}). As such, they conjecturally exhibit certain features, which demand testing (and eventually proofs): the $N$-transform conjecture (Sec.~\ref{sss:RotationalSymmetryAndNTransformConjecture}), which makes them reducible (at least concerning the level density) to Hermitian models; the single ring conjecture (Sec.~\ref{sss:SingleRingConjecture}); the erfc conjecture (Sec.~\ref{sss:ErfcConjecture}). (iii) They contain two operations widely investigated in the literature on random matrices, albeit rather separately: summation (cf.~e.g.~\cite{Stephanov1996,FeinbergZee1997-01,FeinbergZee1997-02,JanikNowakPappWambachZahed1997,JanikNowakPappZahed1997-01,JanikNowakPappZahed1997-02,HaagerupLarsen2000,GorlichJarosz2004,Rogers2010,JaroszNowak2004,JaroszNowak2006}) and multiplication (cf.~e.g.~\cite{GredeskulFreilikher1990,CrisantiPaladinVulpiani1993,Beenakker1997,Caswell2000,JacksonLautrupJohansenNielsen2002,JanikWieczorek2003,GudowskaNowakJanikJurkiewiczNowak20032005,TulinoVerdu2004,NarayananNeuberger2007,BanicaBelinschiCapitaineCollins2007,BlaizotNowak2008,LohmayerNeubergerWettig2008,BenaychGeorges2008,KanzieperSingh2010,BurdaJanikWaclaw2009,BurdaJaroszLivanNowakSwiech20102011,Jarosz2010-01,PensonZyczkowski2011,Rogers2010}).


\subsubsection{Applications to quantum information theory}
\label{sss:ApplicationsToQuantumInformationTheory}

The models $\mathbf{T}$ and $\mathbf{W}$ are encountered in the theory of quantum information. This application has been described in~\cite{ZyczkowskiPensonNechitaCollins2010,SEMZyczkowski2010}, as well as in Sec.~I B 2 of~\cite{Jarosz2011-01}, and I refer the reader to these sources for a more detailed exposition (the textbook~\cite{BengtssonZyczkowski2006} is an introduction to the subject). It is debatable whether one can find the model $\mathbf{V}$ in this theory; if not, the reader may treat it just as a mathematically natural generalization of $\mathbf{W}$.

\emph{Density matrix.} Fundamental objects in quantum information theory are ``mixed states,'' i.e., statistical ensembles of quantum states. Such random states arise in a number of important settings: (i) If a quantum system interacts with its environment in a complicated (noisy) way, which may be regarded as random. (ii) If a quantum system is in thermal equilibrium. (iii) If the preparation history of a quantum system is unknown or uncertain (such as for quantum analogues of classically chaotic systems). (iv) If one investigates generic properties of an unknown complicated quantum system, one may assume it is random. (v) If a system consists of subsystems which are entangled, each of them must be described by a mixed state (and quantum entanglement is a central feature in the theory of quantum computers). A classic example is light polarization: a polarized photon can be written as a superposition of two helicities, right and left circular polarizations, $( a | R \rangle + b | L \rangle )$ (a ``pure state''); whereas unpolarized light may be described as being $| R \rangle$ or $| L \rangle$, each with probability $1 / 2$ (a mixed state).

A mixed state cannot be represented by a single state vector---a proper formalism is that of a ``density matrix.'' If one considers a statistical mixture of $N$ pure states \smash{$| \psi_{i} \rangle$}, each with probability \smash{$p_{i} \in [ 0 , 1 ]$} (\smash{$\sum_{i = 1}^{N} p_{i} = 1$}), the density matrix is defined as \smash{$\boldsymbol{\rho} \equiv \sum_{i = 1}^{N} p_{i} | \psi_{i} \rangle \langle \psi_{i} |$} (a convex combination of pure states; called sometimes an ``incoherent superposition''), and it has this genuine property that the expectation value of any observable $\mathbf{A}$ is given by $\Tr ( \boldsymbol{\rho} \mathbf{A} )$. More generally, any operator which is Hermitian, positive-semidefinite (its eigenvalues are nonnegative) and has trace one (its eigenvalues sum to one) may be considered a density matrix.

I will be interested in complicated composite quantum systems. For example, for a bi-partite system consisting of a subsystem $\mathcal{A}$ of size \smash{$N_{1}$} and $\mathcal{B}$ of size \smash{$N_{2}$} (with orthonormal bases, \smash{$\{ | i \rangle_{\mathcal{A}} \}$} and \smash{$\{ | j \rangle_{\mathcal{B}} \}$}), a general pure state of the full system is ``entangled,'' i.e., it cannot be written as a tensor product of pure states of $\mathcal{A}$ and $\mathcal{B}$,
\begin{equation}\label{eq:EntangledStateOfABipartiteSystem}
| \psi \rangle \equiv \sum_{i = 1}^{N_{1}} \sum_{j = 1}^{N_{2}} X_{i j} | i \rangle_{\mathcal{A}} \otimes | j \rangle_{\mathcal{B}} .
\end{equation}
In other words, $\mathcal{A}$ or $\mathcal{B}$ cannot be said to be in any definite pure state. But they can be characterized by a density matrix---consider any observable $\mathbf{A}$ on $\mathcal{A}$; its expectation value in the state $| \psi \rangle$ reads \smash{$\Tr ( \boldsymbol{\rho}_{\mathcal{A}} \mathbf{A} )$}, where the ``reduced density matrix'' of $\mathcal{A}$,
\begin{equation}\label{eq:ReducedDensityMatrix}
\boldsymbol{\rho}_{\mathcal{A}} \equiv \frac{\mathbf{X} \mathbf{X}^{\dagger}}{\Tr \left( \mathbf{X} \mathbf{X}^{\dagger} \right)} ,
\end{equation}
and $\mathbf{X}$ is rectangular \smash{$N_{1} \times N_{2}$} complex matrix. (This may equivalently be obtained by performing the ``partial trace'' \smash{$\Tr_{\mathcal{B}}$} of the density matrix of the full system, $| \psi \rangle \langle \psi |$, i.e., \smash{$\sum_{j} {}_{\mathcal{B}} \langle j | \psi \rangle \langle \psi | j \rangle_{\mathcal{B}}$}, and normalizing it.) Therefore, the eigenvalues of \smash{$\boldsymbol{\rho}_{\mathcal{A}}$} (which are the properly normalized singular values of $\mathbf{X}$) are precisely the above probabilities \smash{$p_{i}$}. This picture should be supplied by the discussed above basic notion of replacing the complicatedness of the full system by randomness---considering $\mathbf{X}$ to be some random matrix, and calculating the mean density of its singular values.

An important measure for a mixed state is its ``von Neumann entropy,''
\begin{equation}\label{eq:VonNeumannEntropy}
\mathcal{S} ( \boldsymbol{\rho} ) \equiv - \Tr ( \boldsymbol{\rho} \log \boldsymbol{\rho} ) = - \sum_{i = 1}^{N} p_{i} \log p_{i} .
\end{equation}
It represents the degree of randomness (mixture) in the mixed state (thus, quantum information; thus, it also measures entanglement)---it is the larger, the more disperse the probabilities \smash{$p_{i}$} are; it is zero for a pure state, and reaches its maximum $\log N$ for all the probabilities equal [the full system is in the ``maximally entangled (Bell) state'']. Moreover, a measurement can never decrease this entropy; consequently, a measurement may take a pure state to a mixed one, but not conversely (except there is a greater growth of entropy in the environment).

I will focus on some physically motivated ``structured ensembles'' of random states in which an important role is played by the tensor product structure of the Hilbert spaces of the subsystems, i.e., which are invariant under local unitary transformations in the respective Hilbert spaces.

\emph{Model $\mathbf{S}$.} The following construction leads to the appearance of the model $\mathbf{S}$ in the above setting:

(i) Consider a bi-partite system $( \mathcal{A} , \mathcal{B} )$ of size $N \times N$ in the Bell state,
\begin{equation}\label{eq:BellStatePsiPlus}
| \Psi^{+}_{\mathcal{A} \mathcal{B}} \rangle \equiv \frac{1}{\sqrt{N}} \sum_{i = 1}^{N} | i \rangle_{\mathcal{A}} \otimes | i \rangle_{\mathcal{B}} .
\end{equation}

(ii) Consider the following type of randomness---apply to this Bell state $L$ independent random local unitary transformations (belonging to the CUE) \smash{$\mathbf{U}_{l}$} in the principal system $\mathcal{A}$, and form a coherent superposition of the resulting (maximally entangled) states with some weights \smash{$w_{l}$},
\begin{equation}
\begin{split}\label{eq:ModelSInQuantumInformationTheory}
| \psi \rangle &\equiv \sum_{l = 1}^{L} w_{l} \left( \mathbf{U}_{l} \otimes \mathbf{1}_{N} \right) | \Psi^{+}_{\mathcal{A} \mathcal{B}} \rangle =\\
&= \left( \mathbf{S} \otimes \mathbf{1}_{N} \right) | \Psi^{+}_{\mathcal{A} \mathcal{B}} \rangle =\\
&= \frac{1}{\sqrt{N}} \sum_{i , j = 1}^{N} S_{i j} | i \rangle_{\mathcal{A}} \otimes | j \rangle_{\mathcal{B}} .
\end{split}
\end{equation}

(iii) Hence, the reduced density matrix for $\mathcal{A}$ is given by (\ref{eq:ReducedDensityMatrix}) with $\mathbf{X} = \mathbf{S}$. [The normalization constant is \smash{$\Tr ( \mathbf{S} \mathbf{S}^{\dagger} ) = N \sum_{l = 1}^{L} | w_{l} |^{2} + \ldots$}, where the dots are much smaller than $N$ in the large-$N$ limit (\ref{eq:ThermodynamicLimit}). So one may set for convenience, \smash{$\sum_{l = 1}^{L} | w_{l} |^{2} = 1$}, and investigate the singular values of $\mathbf{S}$ rescaled by $N$.]

\emph{Model $\mathbf{T}$.} It is a matter of a nested repetition of the above procedure to obtain the model $\mathbf{T}$. For instance, for $J = 2$:

(i) Consider the above pure state (\ref{eq:ModelSInQuantumInformationTheory}), constructed according to a matrix \smash{$\mathbf{S}_{2}$} of length \smash{$L_{2}$}, and take its \smash{$L_{1}$} copies.

(ii) Perform a random independent CUE rotation \smash{$\mathbf{U}_{1 l}$} of each of these copies, and form their coherent superposition with some weights \smash{$w_{1 l}$}.

(iii) The reduced density matrix for $\mathcal{A}$ (\ref{eq:ReducedDensityMatrix}) will have \smash{$\mathbf{X} = \mathbf{T} = \mathbf{S}_{1} \mathbf{S}_{2}$}. One may continue this process any $J$ times.

\emph{Model $\mathbf{P}$.} This ensemble originates from a measurement in the product basis of Bell states, in the following way:

(i) Consider a composite system with $2 K$ subsystems of sizes
\begin{equation}\label{eq:ModelPInQuantumInformationTheoryDerivation01}
\underbrace{\mathcal{A}_{1} ,}_{\textrm{size } N_{1}} \underbrace{\mathcal{A}_{2} , \mathcal{A}_{3}}_{\textrm{size } N_{2}} , \ldots , \underbrace{\mathcal{A}_{2 K - 2} , \mathcal{A}_{2 K - 1}}_{\textrm{size } N_{K}} , \underbrace{\mathcal{A}_{2 K} .}_{\textrm{size } N_{K + 1}}
\end{equation}

(ii) Take an arbitrary product state \smash{$| \psi_{0} \rangle \equiv | 0 \rangle_{\mathcal{A}_{1}} \otimes | 0 \rangle_{\mathcal{A}_{2}} \otimes \ldots \otimes | 0 \rangle_{\mathcal{A}_{2 K}}$}, and apply to it random unitary local transformations acting on the pairs of subsystems,
\begin{equation}\label{eq:ModelPInQuantumInformationTheoryDerivation02}
| \psi \rangle \equiv \left( \mathcal{U}_{\mathcal{A}_{1} \mathcal{A}_{2}} \otimes \mathcal{U}_{\mathcal{A}_{3} \mathcal{A}_{4}} \otimes \ldots \otimes \mathcal{U}_{\mathcal{A}_{2 K - 1} \mathcal{A}_{2 K}} \right) | \psi_{0} \rangle .
\end{equation}
The result is separable with respect to the above pairing, i.e., it can be expanded in the product basis as
\begin{equation}
\begin{split}\label{eq:ModelPInQuantumInformationTheoryDerivation03}
| \psi \rangle = &\sum_{i_{1} = 1}^{N_{1}} \sum_{i_{2} , i_{2}^{\prime} = 1}^{N_{2}} \ldots \sum_{i_{K} , i_{K}^{\prime} = 1}^{N_{K}} \sum_{i_{K + 1} = 1}^{N_{K + 1}}\\
&[ \mathbf{A}_{1} ]_{i_{1} i_{2}} [ \mathbf{A}_{2} ]_{i_{2}^{\prime} i_{3}} \ldots [ \mathbf{A}_{K - 1} ]_{i_{K - 1}^{\prime} i_{K}} [ \mathbf{A}_{K} ]_{i_{K} i_{K + 1}} \cdot\\
&\cdot | i_{1} \rangle_{\mathcal{A}_{1}} \otimes | i_{2} \rangle_{\mathcal{A}_{2}} \otimes | i_{2}^{\prime} \rangle_{\mathcal{A}_{3}} \otimes \ldots \otimes | i_{K + 1} \rangle_{\mathcal{A}_{2 K}} ,
\end{split}
\end{equation}
where the coefficients, gathered into $K$ rectangular (\smash{$N_{k} \times N_{k + 1}$}) matrices \smash{$\mathbf{A}_{k}$}, may be assumed independent Gaussian (\ref{eq:RectangularGinUEJPDF}).

(iii) Consider the Bell states (\ref{eq:BellStatePsiPlus}) on the pairs \smash{$( \mathcal{A}_{2} , \mathcal{A}_{3} )$}, \ldots, \smash{$( \mathcal{A}_{2 K - 2} , \mathcal{A}_{2 K - 1} )$}, and project $| \psi \rangle$ onto their product,
\begin{equation}\label{eq:ModelPInQuantumInformationTheoryDerivation04}
\mathcal{P} \equiv \mathbf{1}_{\mathcal{A}_{1}} \otimes \left( \bigotimes_{k = 2}^{K} | \Psi^{+}_{\mathcal{A}_{2 k - 2} \mathcal{A}_{2 k - 1}} \rangle \langle \Psi^{+}_{\mathcal{A}_{2 k - 2} \mathcal{A}_{2 k - 1}} | \right) \otimes \mathbf{1}_{\mathcal{A}_{2 K}} ,
\end{equation}
which leads to a random pure state describing the remaining two subsystems, \smash{$\mathcal{A}_{1}$} and \smash{$\mathcal{A}_{2 K}$}, through the matrix $\mathbf{P}$,
\begin{equation}
\begin{split}\label{eq:ModelPInQuantumInformationTheoryDerivation05}
| \phi \rangle &\equiv \mathcal{P} | \psi \rangle \propto\\
&\propto \sum_{i_{1} = 1}^{N_{1}} \sum_{i_{K + 1} = 1}^{N_{K + 1}} [ \mathbf{P} ]_{i_{1} i_{K + 1}} | i_{1} \rangle_{\mathcal{A}_{1}} \otimes | i_{K + 1} \rangle_{\mathcal{A}_{2 K}} .
\end{split}
\end{equation}

(iv) The reduced density matrix for \smash{$\mathcal{A}_{1}$} (i.e., the normalized partial trace over \smash{$\mathcal{A}_{2 K}$}) is therefore (\ref{eq:ReducedDensityMatrix}) with $\mathbf{X} = \mathbf{P}$.

\emph{Model $\mathbf{W}$.} A direct combination of the above two algorithms leads to the model $\mathbf{W}$---one should consider the random pure states corresponding to the model $\mathbf{T}$ on the $2 K$ subsystems (\ref{eq:ModelPInQuantumInformationTheoryDerivation01}), and proceed as above.

One may find in the literature only an expression for the mean density of the singular values of $\mathbf{W}$ in the case of $J = 1$, \smash{$L_{1} = 2$}, \smash{$w_{1 l} = 1 / \sqrt{2}$} (for $l = 1 , 2$), $K = 1$, \smash{$\sigma_{1} = 1$}, \smash{$r_{1} = 1$}---the ``Bures distribution''~\cite{Bures1969,SommersZyczkowski2004},
\begin{equation}
\begin{split}\label{eq:BuresDistribution}
\rho_{\mathbf{W}^{\dagger} \mathbf{W}} ( x ) = &\frac{1}{4 \sqrt{3} \pi} \Bigg( \left( \frac{\beta}{x} + \sqrt{\frac{\beta^{2}}{x^{2}} - 1} \right)^{2 / 3} -\\
&-\left( \frac{\beta}{x} - \sqrt{\frac{\beta^{2}}{x^{2}} - 1} \right)^{2 / 3} \Bigg) ,
\end{split}
\end{equation}
for $x \in [ 0 , \beta ]$ and zero otherwise, where for short, \smash{$\beta \equiv 3 \sqrt{3}$}. The chief purpose of this paper is to extend this result to arbitrary values of the parameters, even to a product of a number of matrices $\mathbf{W}$, as well as to the mean density of the eigenvalues.


\section{Generalized Bures products}
\label{s:GeneralizedBuresProducts}


\subsection{Generalized multiplication laws}
\label{ss:GeneralizedMultiplicationLaws}

This paper deals with products of random rectangular or non-Hermitian matrices, hence, I will start from adjusting the free probability multiplication law (\ref{eq:MultiplicationLaw}) to such a situation. Consider a product of arbitrary independent random matrices,
\begin{equation}\label{eq:XDefinition}
\mathbf{X} \equiv \mathbf{X}_{1} \mathbf{X}_{2} \ldots \mathbf{X}_{I} ,
\end{equation}
where \smash{$\mathbf{X}_{i}$}, $i = 1 , 2 , \ldots , I$, is rectangular of dimensions \smash{$T_{i} \times T_{i + 1}$}, which tend to infinity in such a way that the rectangularity ratios
\begin{equation}\label{eq:sDefinition}
s_{i} \equiv \frac{T_{i}}{T_{I + 1}} ,
\end{equation}
remain finite. I will follow the steps sketched in~Sec.~\ref{sss:FreeProbabilityAndModelP} (cf.~\cite{BurdaJaroszLivanNowakSwiech20102011}) to derive the mean densities of its singular values and if \smash{$s_{1} = 1$} also the eigenvalues.


\subsubsection{Singular values of the product $\mathbf{X}$ via cyclic shifts and multiplication law}
\label{sss:SingularValuesOfTheProductXViaCyclicShiftsAndMultiplicationLaw}

Let me for simplicity set $I = 2$ here. I begin from the eigenvalues of the Hermitian \smash{$T_{3} \times T_{3}$} random matrix,
\begin{equation}\label{eq:SingularValuesOfTheProductXViaCyclicShiftsAndMultiplicationLawDerivation01}
\mathbf{X}^{\dagger} \mathbf{X} = \mathbf{X}_{2}^{\dagger} \mathbf{X}_{1}^{\dagger} \mathbf{X}_{1} \mathbf{X}_{2} .
\end{equation}
However, as a first step, consider instead the \smash{$T_{2} \times T_{2}$} matrix,
\begin{equation}\label{eq:SingularValuesOfTheProductXViaCyclicShiftsAndMultiplicationLawDerivation02}
\mathbf{Y} \equiv \left( \mathbf{X}_{1}^{\dagger} \mathbf{X}_{1} \right) \left( \mathbf{X}_{2} \mathbf{X}_{2}^{\dagger} \right) ,
\end{equation}
which differs from the previous one only by a cyclic shift of the terms. Therefore, as follows from (\ref{eq:HolomorphicMTransformMomentExpansion}), their $N$-transforms are related by
\begin{equation}\label{eq:SingularValuesOfTheProductXViaCyclicShiftsAndMultiplicationLawDerivation03}
N_{\mathbf{X}^{\dagger} \mathbf{X}} ( z ) = N_{\mathbf{Y}} \left( \frac{T_{3}}{T_{2}} z \right) .
\end{equation}
Now, $\mathbf{Y}$ is a product of two free Hermitian random matrices (their freeness is what I actually mean by the assumed ``independence'' of \smash{$\mathbf{X}_{1}$} and \smash{$\mathbf{X}_{2}$}), and thus the multiplication law (\ref{eq:MultiplicationLaw}) can be applied,
\begin{equation}
\begin{split}\label{eq:SingularValuesOfTheProductXViaCyclicShiftsAndMultiplicationLawDerivation04}
N_{\mathbf{Y}} ( z ) &= \frac{z}{z + 1} N_{\mathbf{X}_{1}^{\dagger} \mathbf{X}_{1}} ( z ) N_{\mathbf{X}_{2} \mathbf{X}_{2}^{\dagger}} ( z ) =\\
&= \frac{z}{z + 1} N_{\mathbf{X}_{1}^{\dagger} \mathbf{X}_{1}} ( z ) N_{\mathbf{X}_{2}^{\dagger} \mathbf{X}_{2}} \left( \frac{T_{2}}{T_{3}} z \right) ,
\end{split}
\end{equation}
where in the last step I used an analogue of (\ref{eq:SingularValuesOfTheProductXViaCyclicShiftsAndMultiplicationLawDerivation03}). Finally, inserting (\ref{eq:SingularValuesOfTheProductXViaCyclicShiftsAndMultiplicationLawDerivation04}) into (\ref{eq:SingularValuesOfTheProductXViaCyclicShiftsAndMultiplicationLawDerivation03}), one obtains a multiplication law which allows to calculate the singular values of $\mathbf{X}$ from the singular values of \smash{$\mathbf{X}_{1}$} and \smash{$\mathbf{X}_{2}$},
\begin{equation}\label{eq:SingularValuesOfTheProductXViaCyclicShiftsAndMultiplicationLawDerivation05}
N_{\mathbf{X}^{\dagger} \mathbf{X}} ( z ) = \frac{z}{z + \frac{T_{2}}{T_{3}}} N_{\mathbf{X}_{1}^{\dagger} \mathbf{X}_{1}} \left( \frac{T_{3}}{T_{2}} z \right) N_{\mathbf{X}_{2}^{\dagger} \mathbf{X}_{2}} ( z ) .
\end{equation}

This formula may be generalized to any $I$,
\begin{equation}\label{eq:SingularValuesOfTheProductXViaCyclicShiftsAndMultiplicationLawDerivation06}
N_{\mathbf{X}^{\dagger} \mathbf{X}} ( z ) = \frac{z^{I - 1}}{\prod_{i = 2}^{I} \left( z + s_{i} \right)} \prod_{i = 1}^{I} N_{\mathbf{X}_{i}^{\dagger} \mathbf{X}_{i}} \left( \frac{z}{s_{i + 1}} \right)
\end{equation}
[cf.~Eq.~(58) of the first position in~\cite{BurdaJaroszLivanNowakSwiech20102011}].


\subsubsection{Eigenvalues of the product $\mathbf{X}$ assuming it is square and has rotationally-symmetric mean spectrum}
\label{sss:EigenvaluesOfTheProductXAssumingItIsSquareAndHasRotationallySymmetricMeanSpectrum}

If \smash{$s_{1} = 1$}, i.e., $\mathbf{X}$ is square, one may also ask about its eigenvalues. Assume that its mean spectrum has the rotational symmetry around zero (\ref{eq:RotationallySymmetricNonHolomorphicMTransformDefinition}). Combining the $N$-transform conjecture (\ref{eq:NTransformConjecture}) with the multiplication law (\ref{eq:SingularValuesOfTheProductXViaCyclicShiftsAndMultiplicationLawDerivation06}), one finds the appropriate formula,
\begin{equation}\label{eq:EigenvaluesOfTheProductXAssumingItIsSquareAndHasRotationallySymmetricMeanSpectrumDerivation01}
\mathfrak{N}_{\mathbf{X}} ( z ) = \prod_{i = 1}^{I} \frac{z}{z + s_{i}} N_{\mathbf{X}_{i}^{\dagger} \mathbf{X}_{i}} \left( \frac{z}{s_{i + 1}} \right) .
\end{equation}


\subsubsection{Eigenvalues of the product $\mathbf{X}$ assuming all \smash{$\mathbf{X}_{i}$} are square and have rotationally-symmetric mean spectra}
\label{sss:EigenvaluesOfTheProductXAssumingAllXiAreSquareAndHaveRotationallySymmetricMeanSpectra}

If moreover all \smash{$s_{i} = 1$}, i.e., all \smash{$\mathbf{X}_{i}$} are square, and also the mean spectra of all \smash{$\mathbf{X}_{i}$} are rotationally symmetric around zero, then the right-hand side of the multiplication law (\ref{eq:EigenvaluesOfTheProductXAssumingItIsSquareAndHasRotationallySymmetricMeanSpectrumDerivation01}) may be expressed through the eigenvalues of \smash{$\mathbf{X}_{i}$} by virtue of the $N$-transform conjecture (\ref{eq:NTransformConjecture}), which yields simply
\begin{equation}\label{eq:EigenvaluesOfTheProductXAssumingAllXiAreSquareAndHaveRotationallySymmetricMeanSpectraDerivation01}
\mathfrak{N}_{\mathbf{X}} ( z ) = \prod_{i = 1}^{I} \mathfrak{N}_{\mathbf{X}_{i}} ( z ) .
\end{equation}

The models $\mathbf{T}$, $\mathbf{W}$, $\mathbf{V}$ all possess the property (\ref{eq:RotationallySymmetricNonHolomorphicMTransformDefinition}) since they are products of terms which separately exhibit this symmetry. Hence, one is allowed to use the multiplication law (\ref{eq:EigenvaluesOfTheProductXAssumingAllXiAreSquareAndHaveRotationallySymmetricMeanSpectraDerivation01}) for them. Eventually, the rotational symmetry will be confirmed by Monte Carlo simulations.


\subsection{Product $\mathbf{T}$}
\label{ss:ProductT}


\subsubsection{Master equations for $\mathbf{T}$}
\label{sss:MasterEquationsForT}

\emph{Eigenvalues.} It is now straightforward to write down the master equations for the nonholomorphic $M$-transform \smash{$\mathfrak{M} \equiv \mathfrak{M}_{\mathbf{T}} ( R^{2} )$} of the product $\mathbf{T}$ (\ref{eq:TDefinition})---they are comprised of the multiplication law (\ref{eq:EigenvaluesOfTheProductXAssumingAllXiAreSquareAndHaveRotationallySymmetricMeanSpectraDerivation01}),
\begin{equation}\label{eq:TMasterEquation0}
\mathfrak{N}_{\mathbf{T}} ( z ) = \prod_{j = 1}^{J} \mathfrak{N}_{\mathbf{S}_{j}} ( z ) ,
\end{equation}
which links the $J$ sets (\ref{eq:SMasterEquation1})-(\ref{eq:SMasterEquation3}),
\begin{subequations}
\begin{align}
z &= \sum_{l = 1}^{L_{j}} M_{j l} ,\label{eq:TMasterEquation1}\\
- C_{j} &= \frac{z ( z + 1 )}{\mathfrak{N}_{\mathbf{S}_{j}} ( z )} ,\label{eq:TMasterEquation2}\\
- C_{j} &= \frac{M_{j l} \left( M_{j l} + 1 \right)}{\left| w_{j l} \right|^{2}} , \quad l = 1 , 2 , \ldots , L_{j} ,\label{eq:TMasterEquation3}
\end{align}
\end{subequations}
for $j = 1 , 2 , \ldots , J$; these are \smash{$( 1 + 2 J + \sum_{j = 1}^{J} L_{j} )$} polynomial equations. They are valid inside the mean spectral domain $\mathcal{D}$, whose borderline (one or two centered circles) is given by (\ref{eq:SRExt})-(\ref{eq:SRInt}).

\emph{Singular values.} These are obtained from the same master equations, albeit with \smash{$\mathfrak{N}_{\mathbf{T}} ( z )$} in (\ref{eq:TMasterEquation0}) replaced by \smash{$\frac{z}{z + 1} N_{\mathbf{T}^{\dagger} \mathbf{T}} ( z )$} (\ref{eq:NTransformConjecture}).

I will now simplify the above equations and solve them either analytically or numerically in a number of special cases, comparing the findings with Monte Carlo simulations.


\subsubsection{Example 0}
\label{sss:TExample0}

\emph{Eigenvalues.} Before that, however, let us remark that if all the terms \smash{$\mathbf{S}_{j}$} have identical lengths, \smash{$L_{j} = L$}, and sequences of weights, \smash{$w_{j l} = w_{l}$}, then (\ref{eq:TMasterEquation1})-(\ref{eq:TMasterEquation3}) imply that all \smash{$\mathfrak{N}_{\mathbf{S}_{j}} ( z )$} are equal to each other. This along with (\ref{eq:TMasterEquation0}) in turn imply that the nonholomorphic $M$-transforms of $\mathbf{T}$ and any \smash{$\mathbf{S}_{j}$} are related by a simple rescaling of the argument,
\begin{equation}\label{eq:TEqualTermsScalingRelation}
\mathfrak{M}_{\mathbf{T}} \left( R^{2} \right) = \mathfrak{M}_{\mathbf{S}} \left( R^{2 / J} \right) .
\end{equation}

\emph{Singular values.} Unfortunately, there is no such scaling relation here; it is clear from trying to follow the above argument in conjunction with (\ref{eq:NTransformConjecture}).


\subsubsection{Example 1}
\label{sss:TExample1}

\begin{figure*}[t]
\includegraphics[width=\columnwidth]{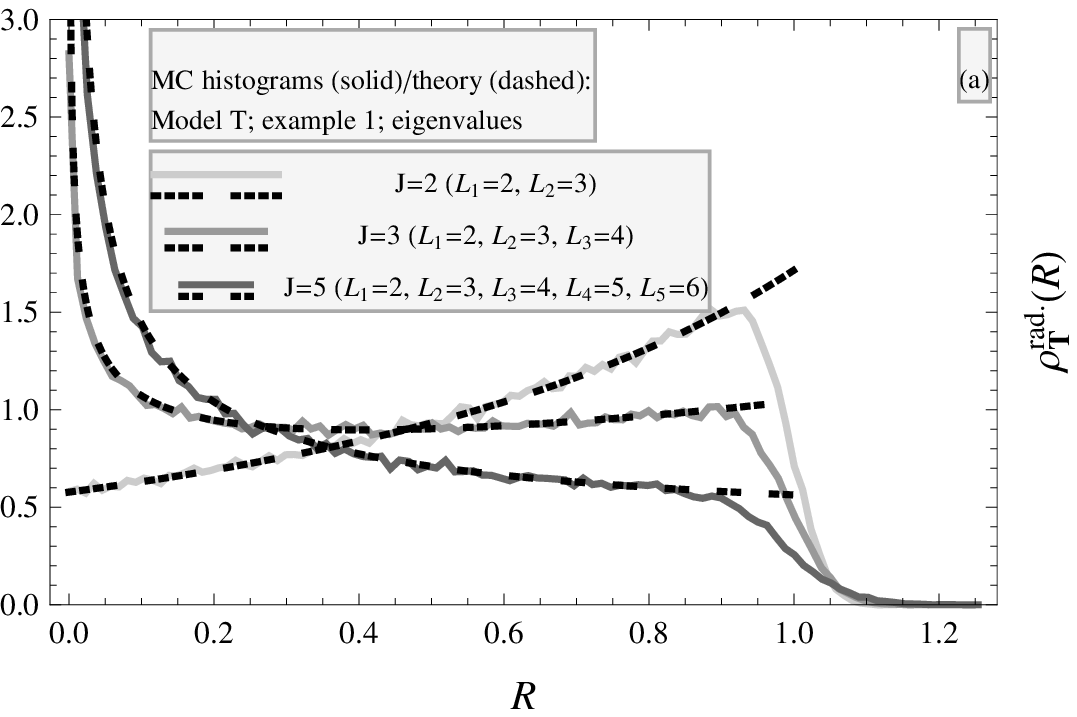}
\includegraphics[width=\columnwidth]{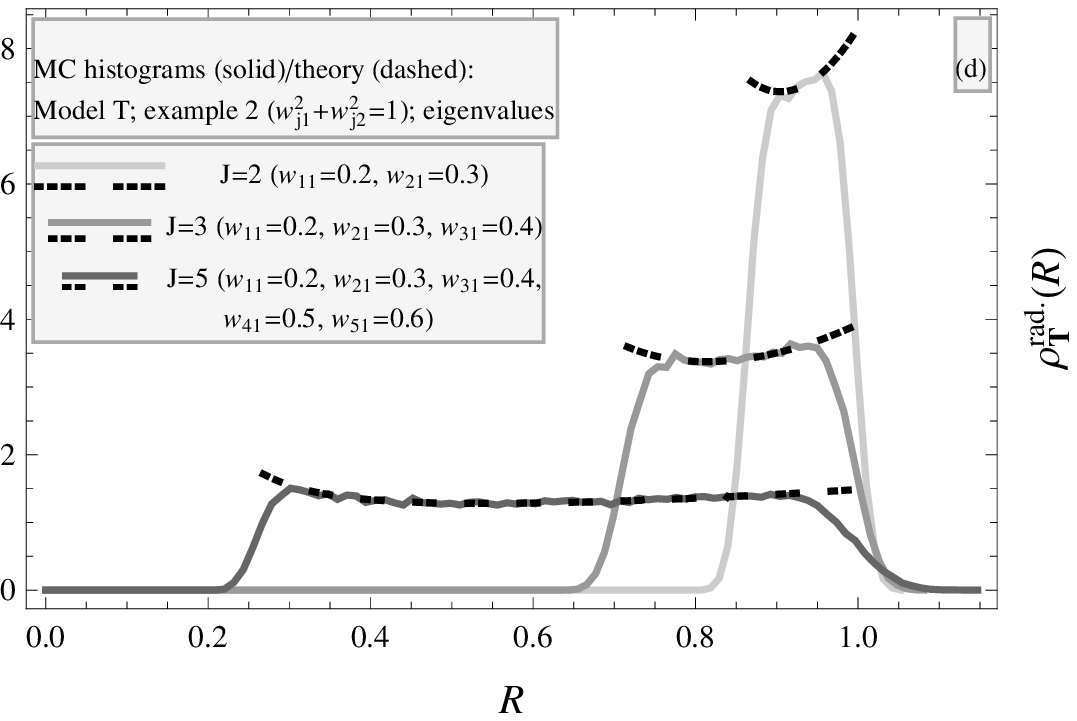}
\includegraphics[width=\columnwidth]{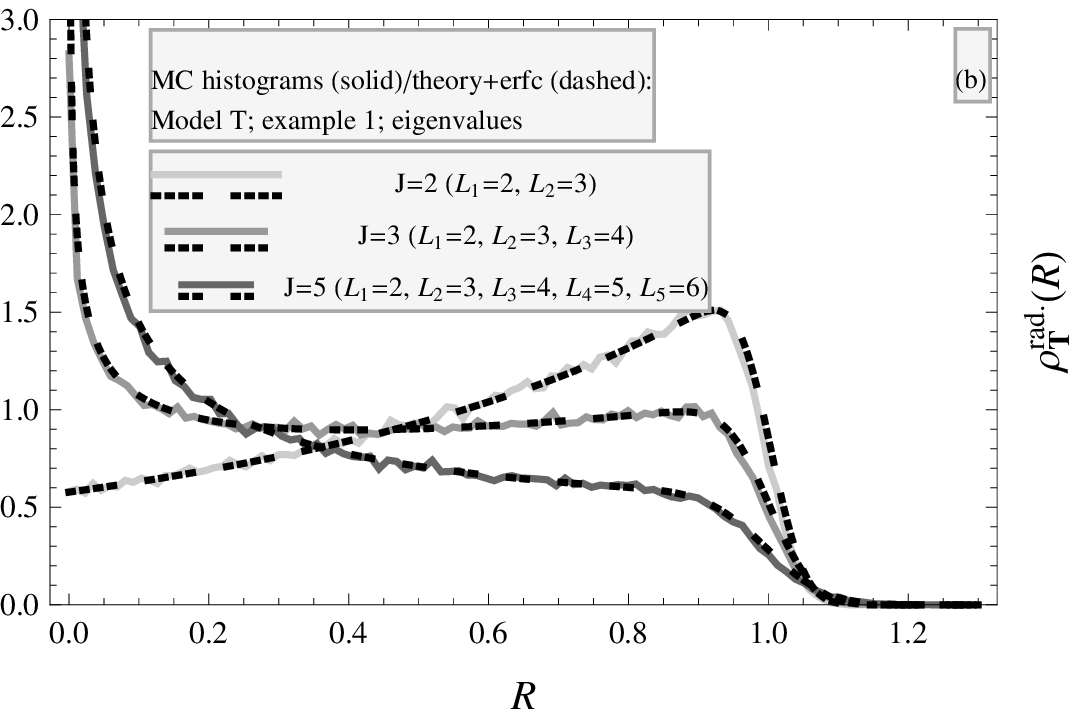}
\includegraphics[width=\columnwidth]{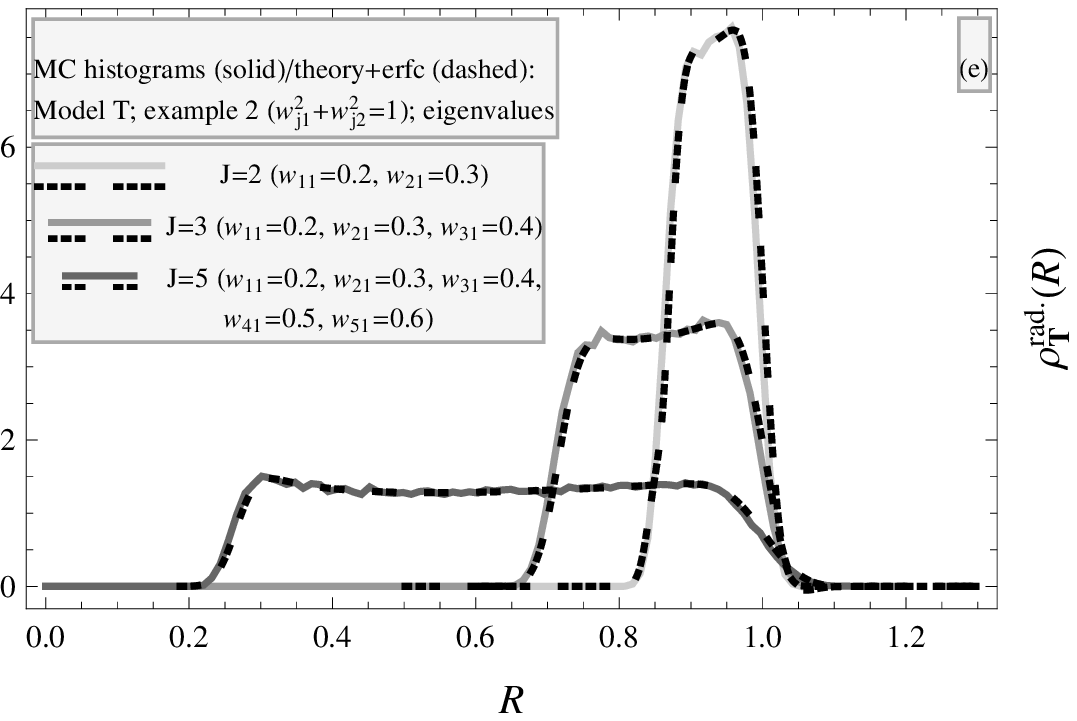}
\includegraphics[width=\columnwidth]{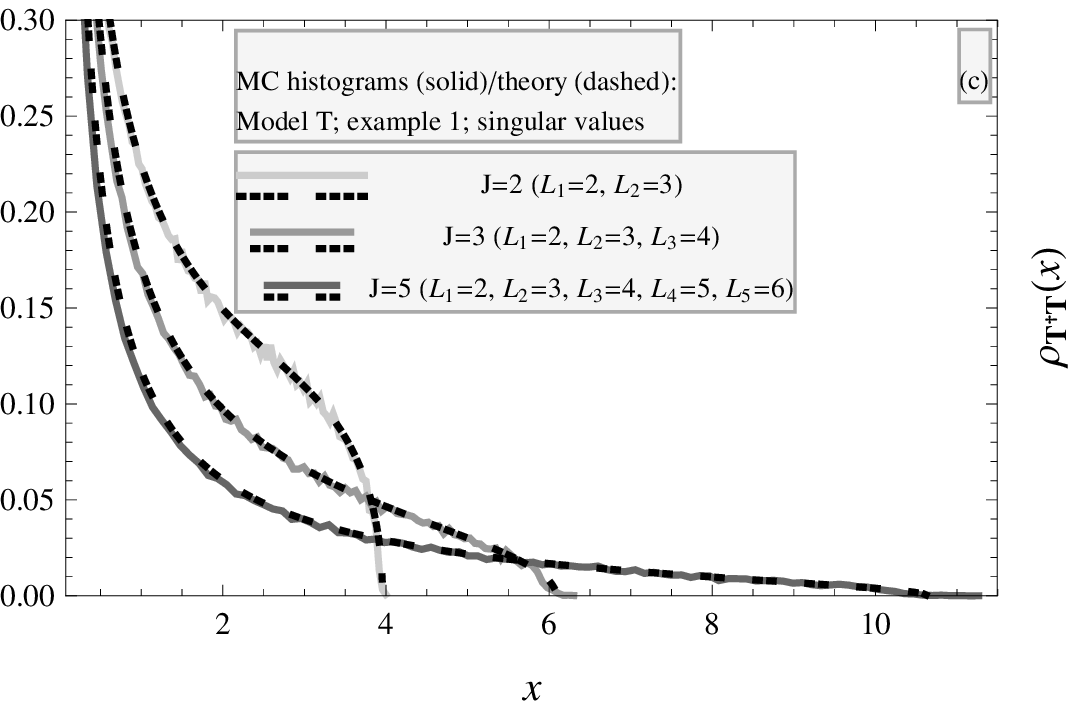}
\includegraphics[width=\columnwidth]{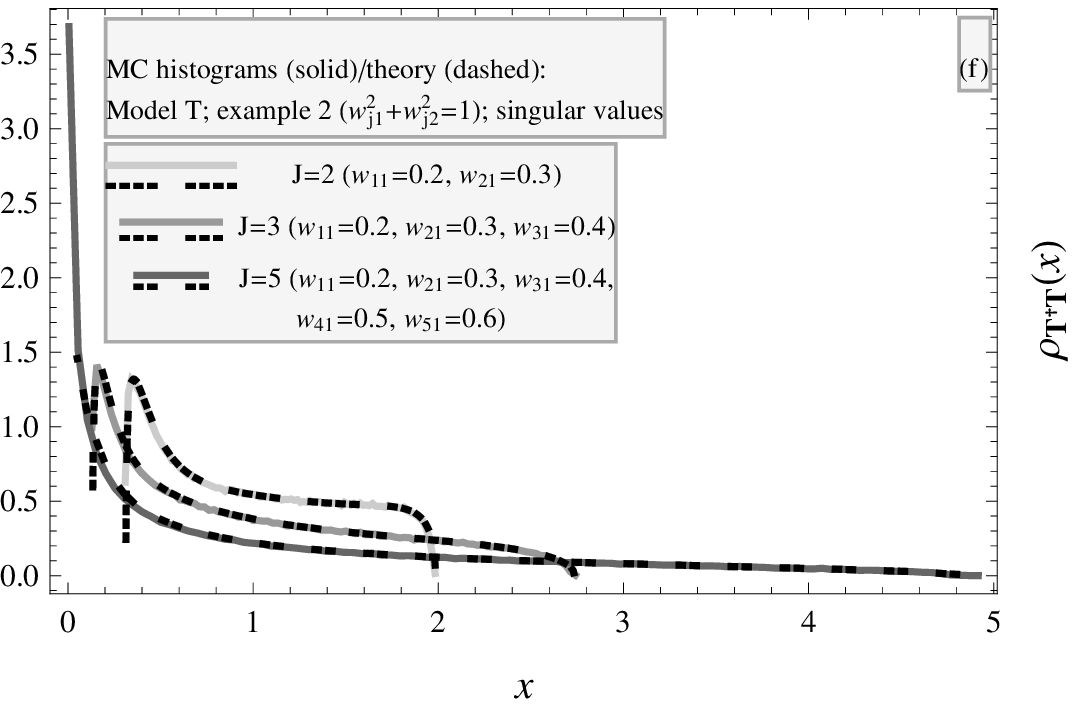}
\caption{Theoretical level densities versus Monte Carlo data for the model $\mathbf{T}$. Top row concerns the eigenvalues, middle row the eigenvalues plus the erfc form-factor, bottom row the singular values. Left column illustrates Example 1, right column Example 2.}
\label{fig:ModelT}
\end{figure*}

\begin{figure}[t]
\includegraphics[width=\columnwidth]{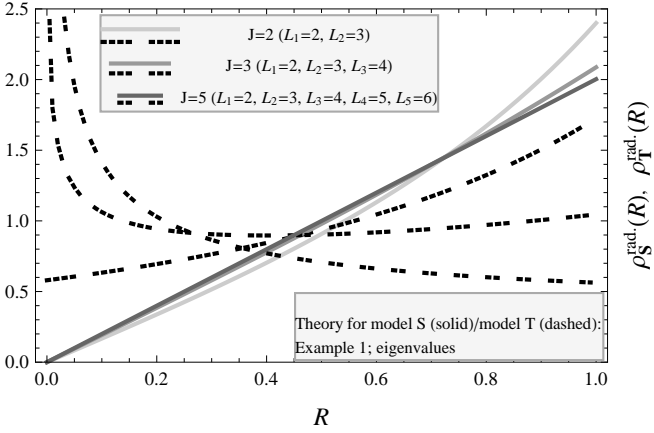}
\caption{Numerical verification of the property described in \emph{Remark 2} in Sec.~\ref{sss:TExample1}, that even though the model $\mathbf{T}$ may be rewritten as a sum of CUE matrices, it is very different from the model $\mathbf{S}$ due to the correlations between the thus obtained CUE terms.}
\label{fig:ModelTCorrelatedU}
\end{figure}

\begin{figure*}[t]
\includegraphics[width=\columnwidth]{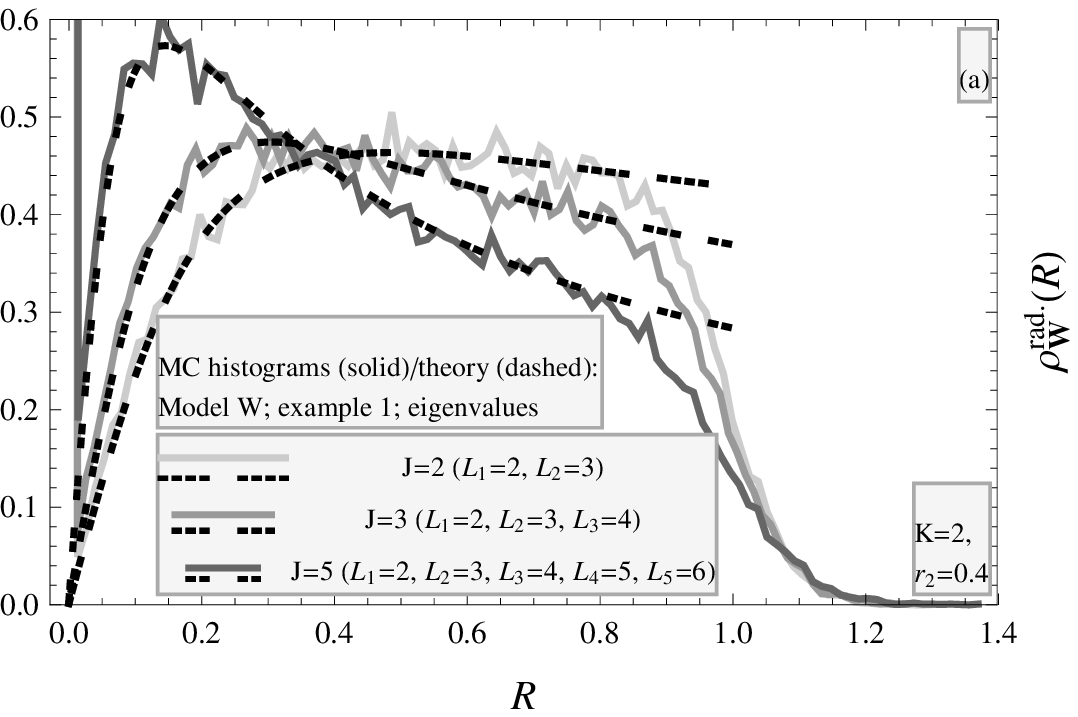}
\includegraphics[width=\columnwidth]{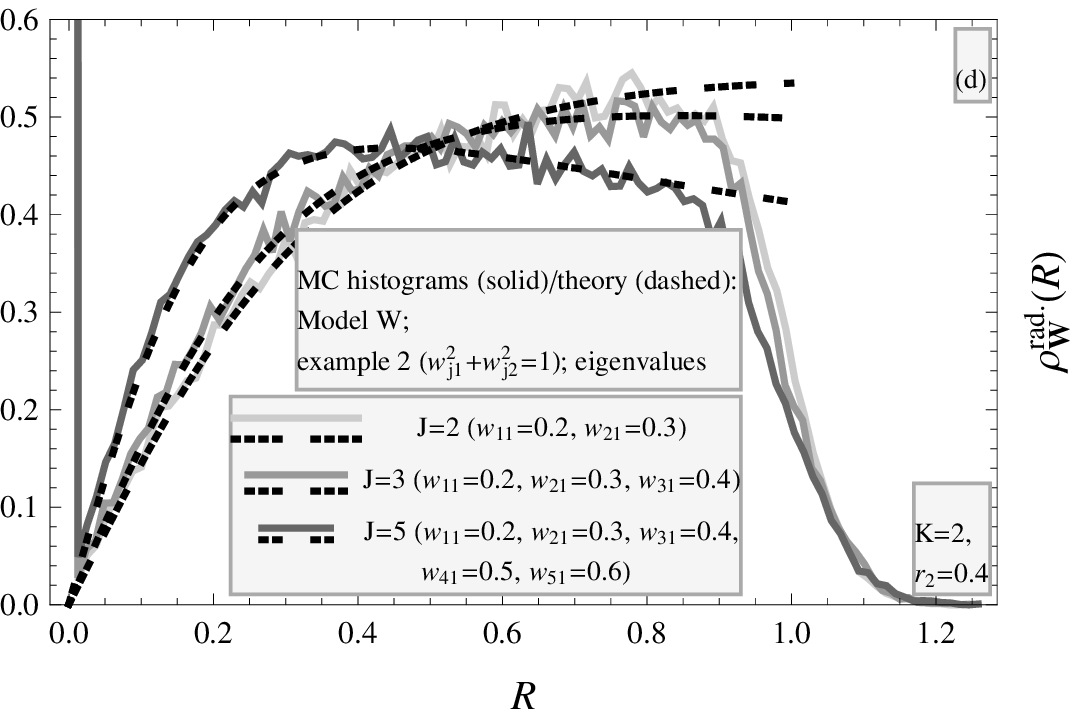}
\includegraphics[width=\columnwidth]{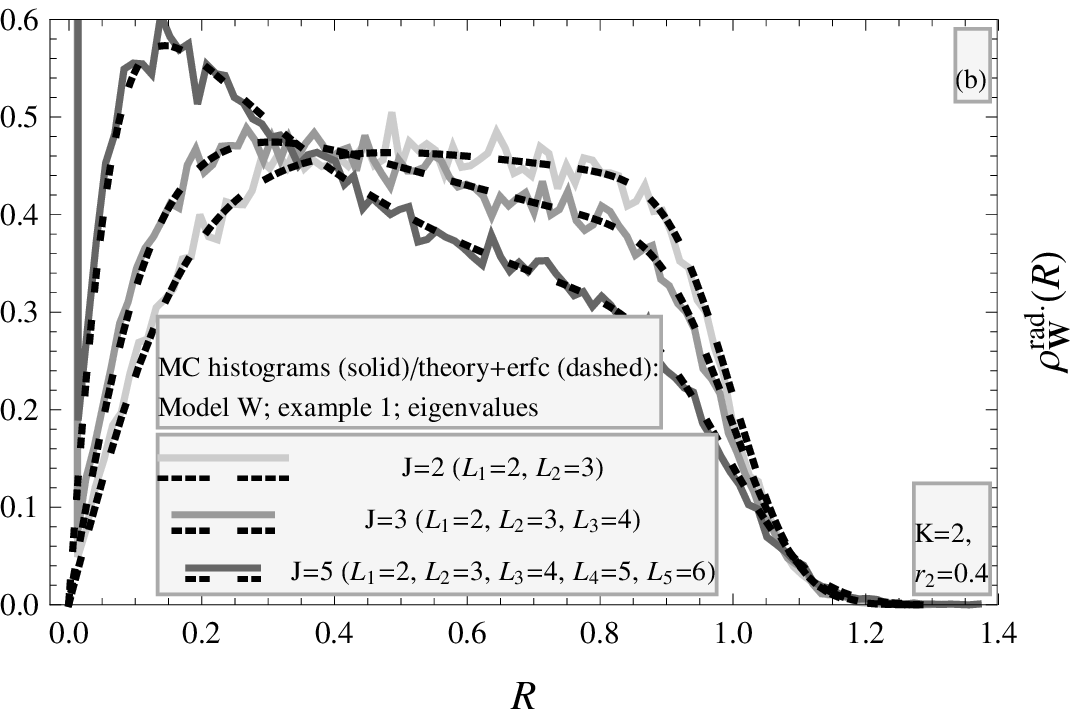}
\includegraphics[width=\columnwidth]{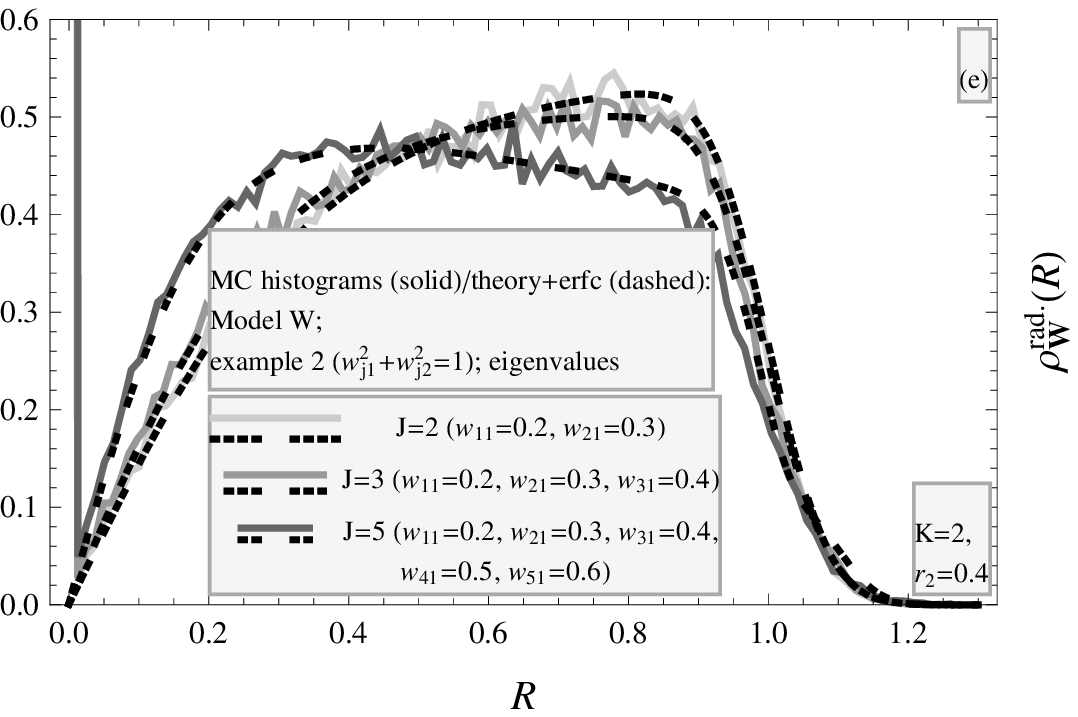}
\includegraphics[width=\columnwidth]{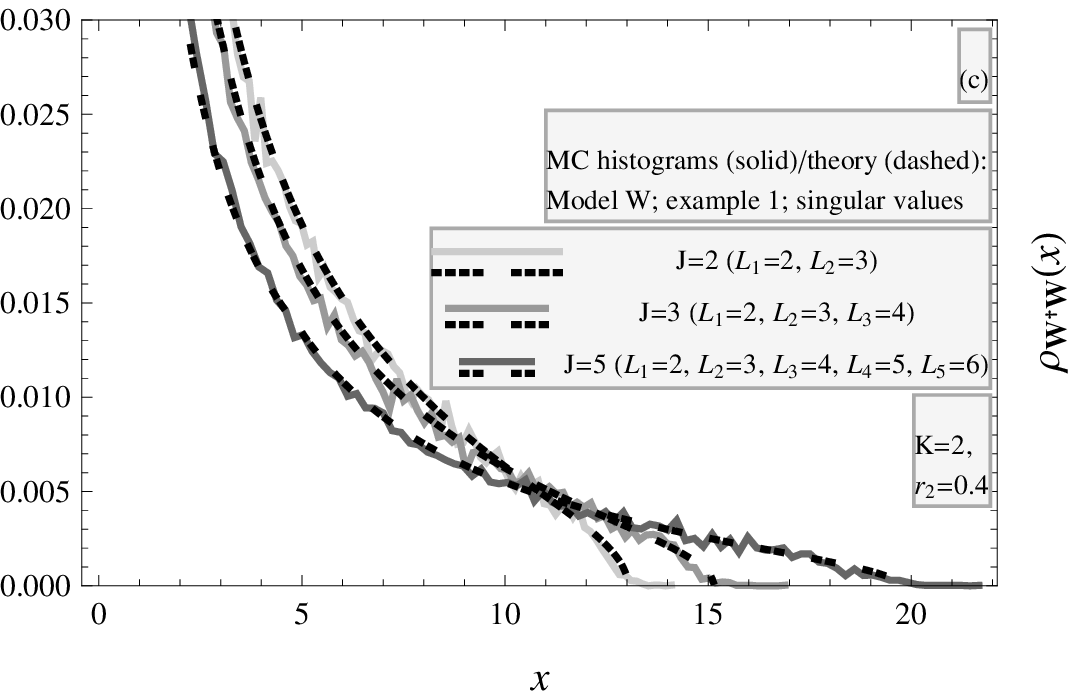}
\includegraphics[width=\columnwidth]{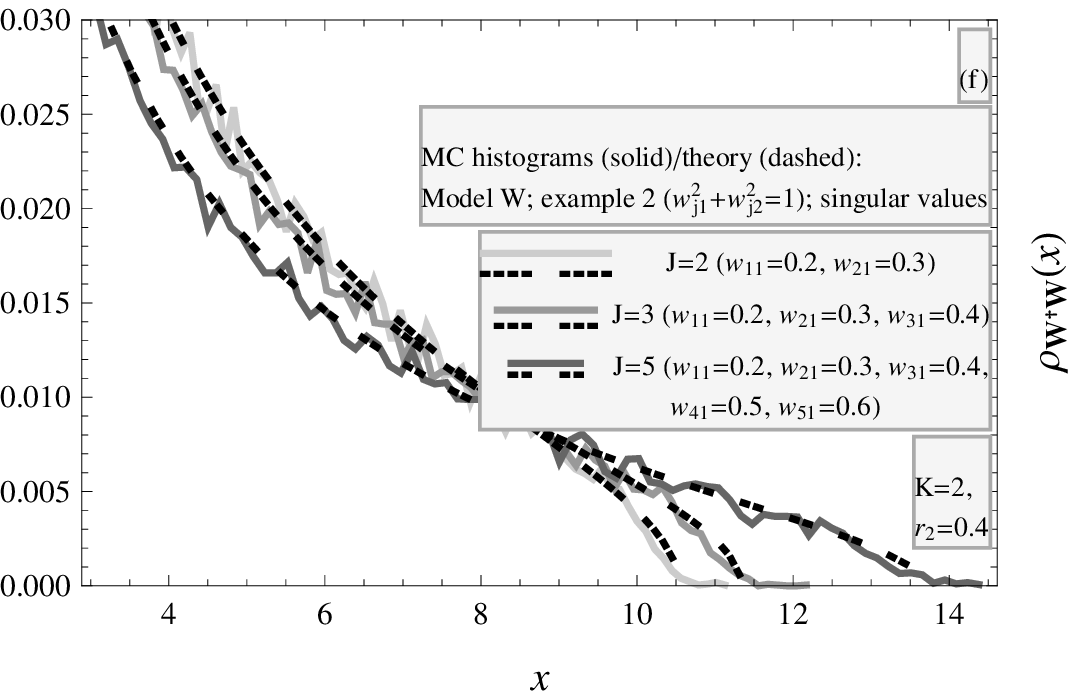}
\caption{Theoretical level densities versus Monte Carlo data for the model $\mathbf{W}$, with $K = 2$. Top row concerns the eigenvalues, middle row the eigenvalues plus the erfc form-factor, bottom row the singular values. Left column illustrates Example 1, right column Example 2.}
\label{fig:ModelWK2}
\end{figure*}

\begin{figure*}[t]
\includegraphics[width=\columnwidth]{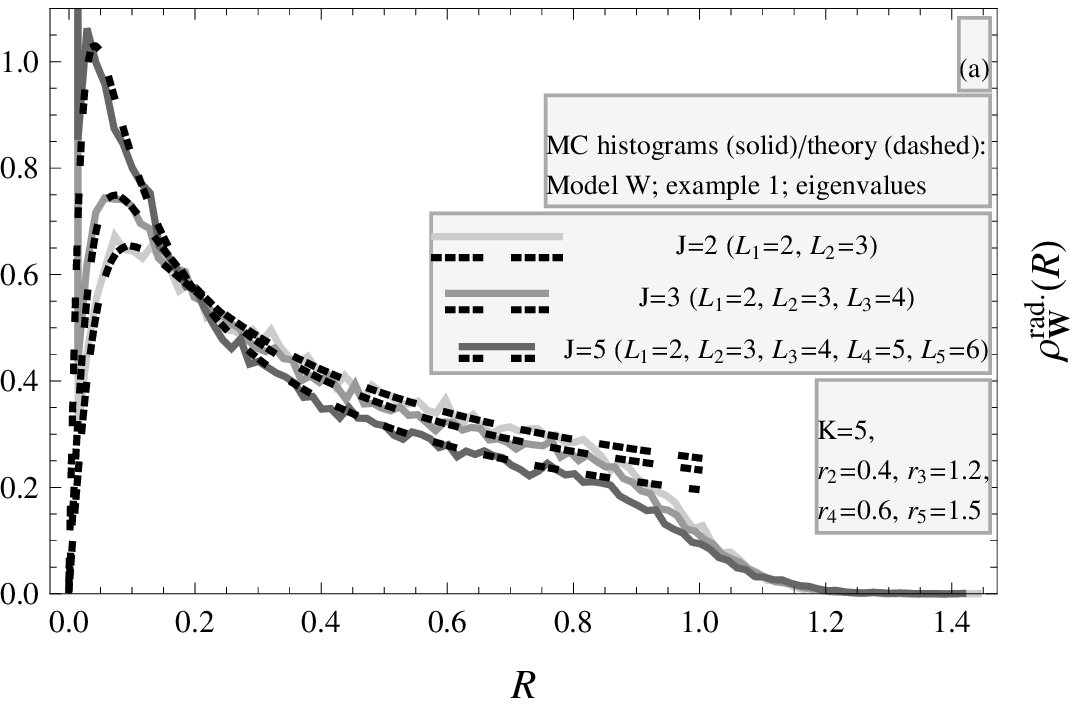}
\includegraphics[width=\columnwidth]{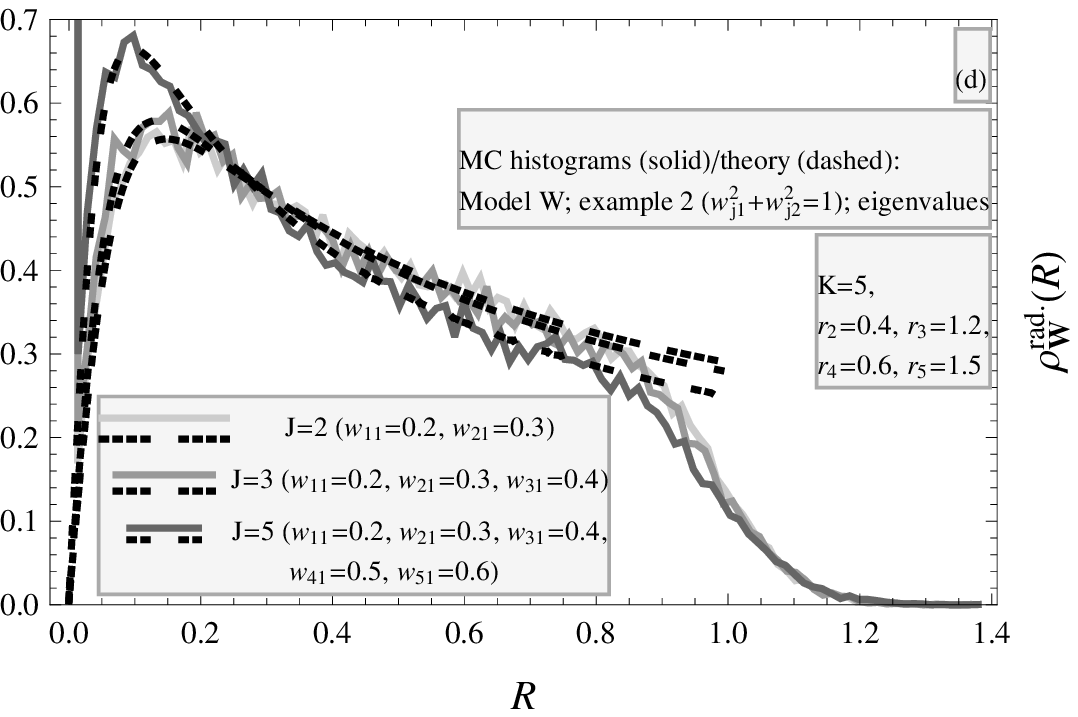}
\includegraphics[width=\columnwidth]{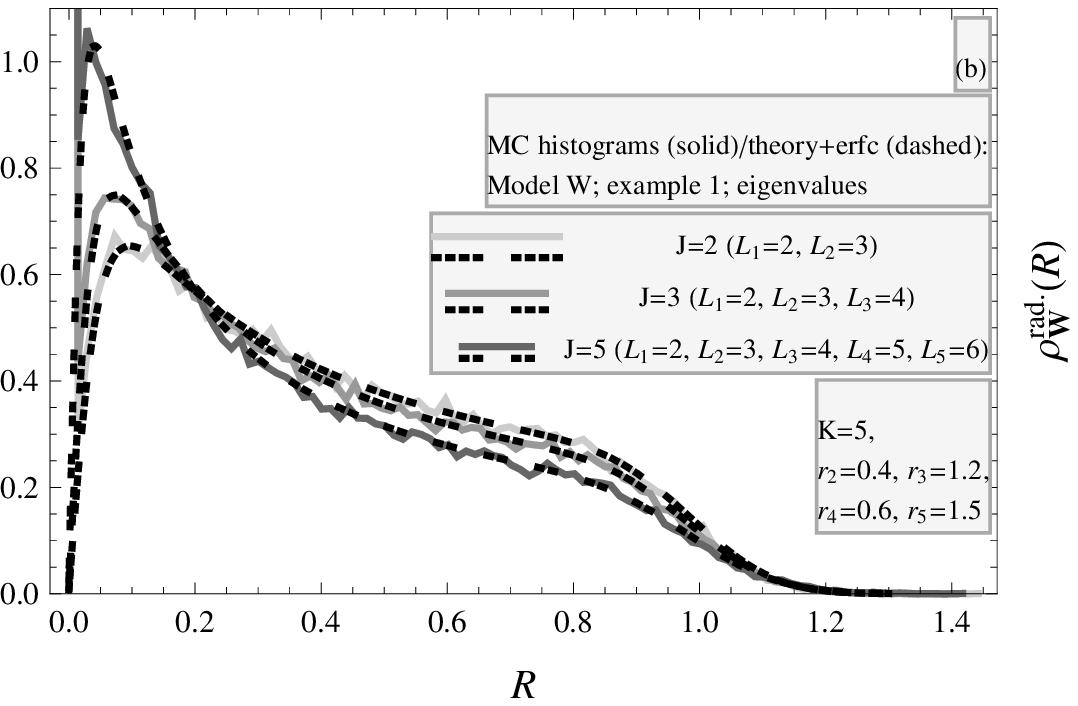}
\includegraphics[width=\columnwidth]{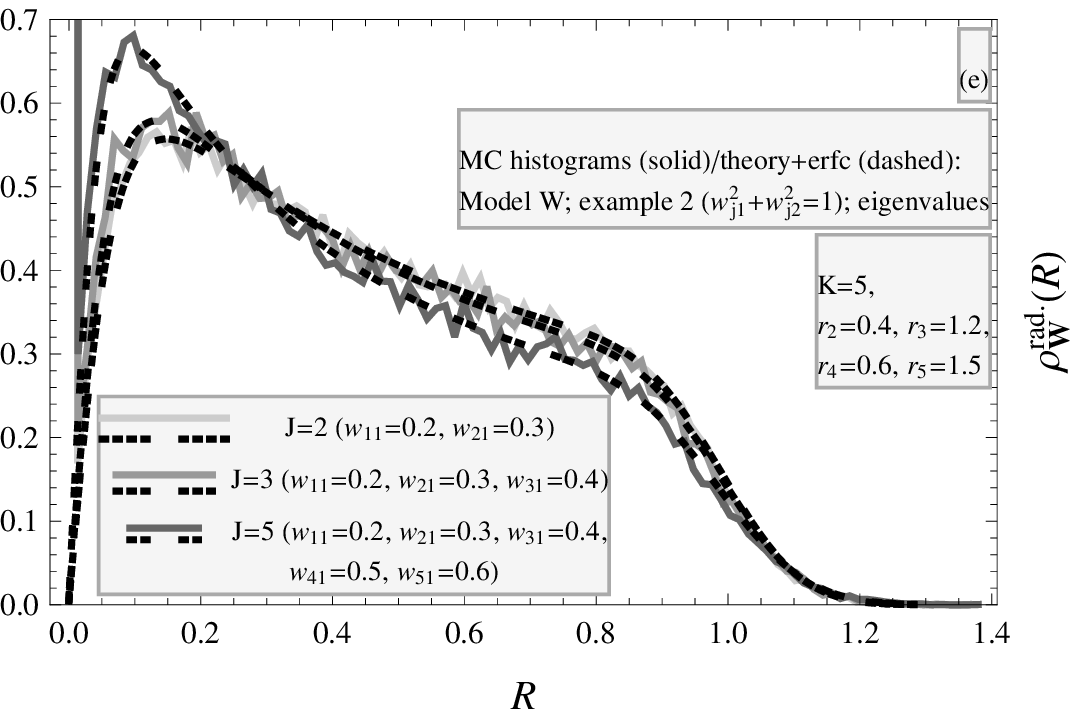}
\includegraphics[width=\columnwidth]{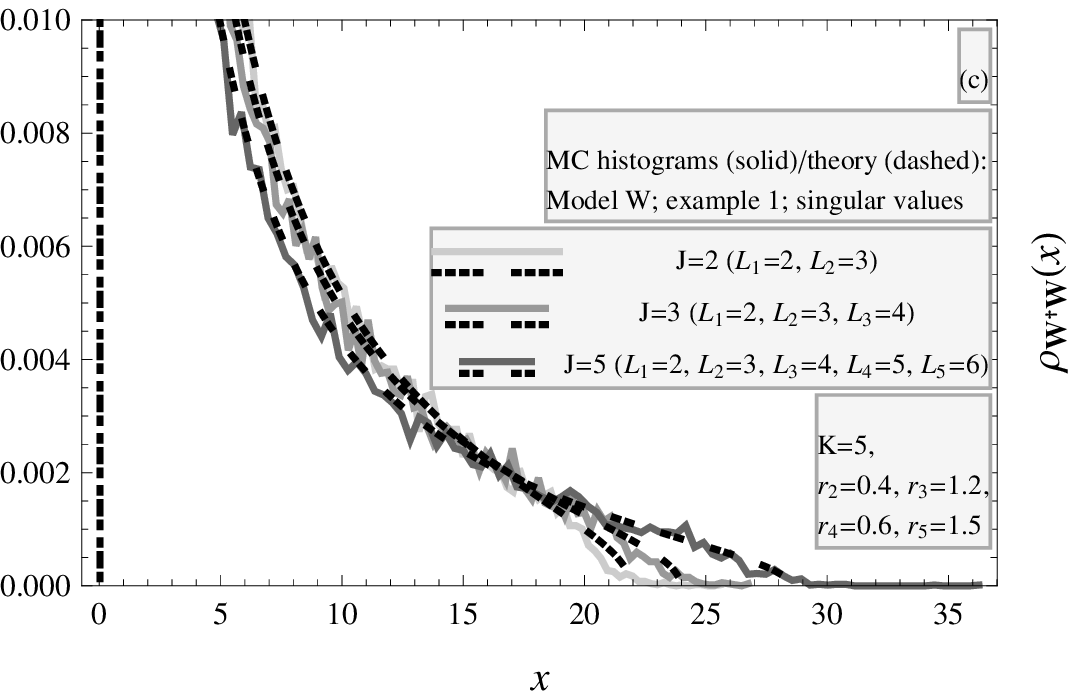}
\includegraphics[width=\columnwidth]{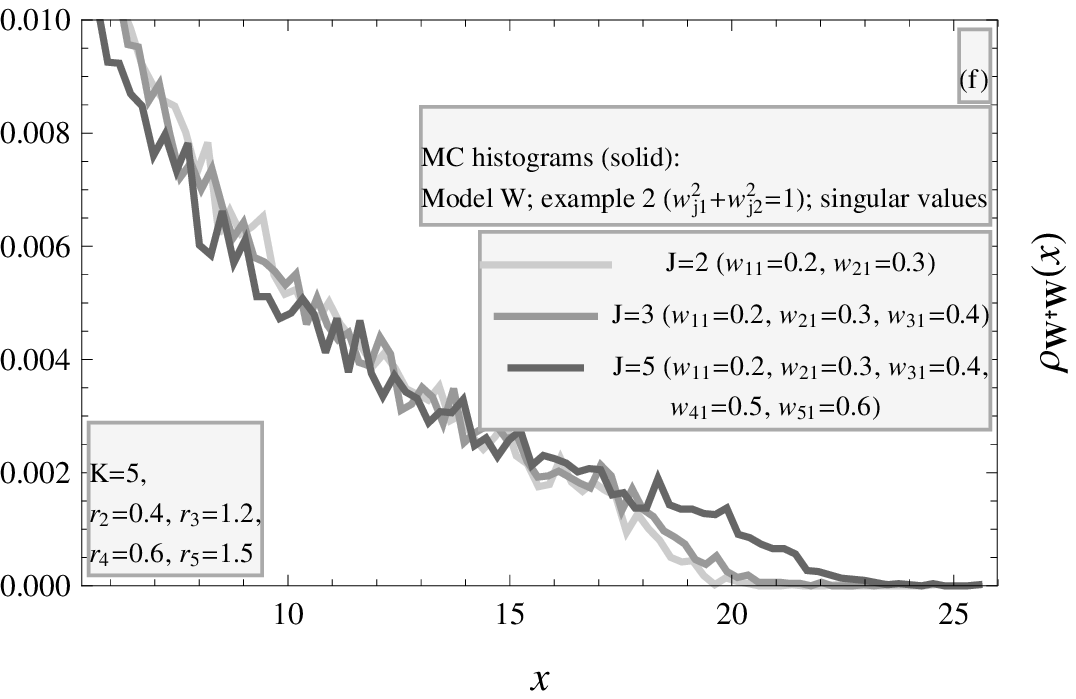}
\caption{Theoretical level densities versus Monte Carlo data for the model $\mathbf{W}$, with $K = 5$. Top row concerns the eigenvalues, middle row the eigenvalues plus the erfc form-factor, bottom row the singular values. Left column illustrates Example 1, right column Example 2. Figure (f) does not include theoretical graphs due to numerical complications encountered in solving the master equations.}
\label{fig:ModelWK5}
\end{figure*}

\begin{figure*}[t]
\includegraphics[width=\columnwidth]{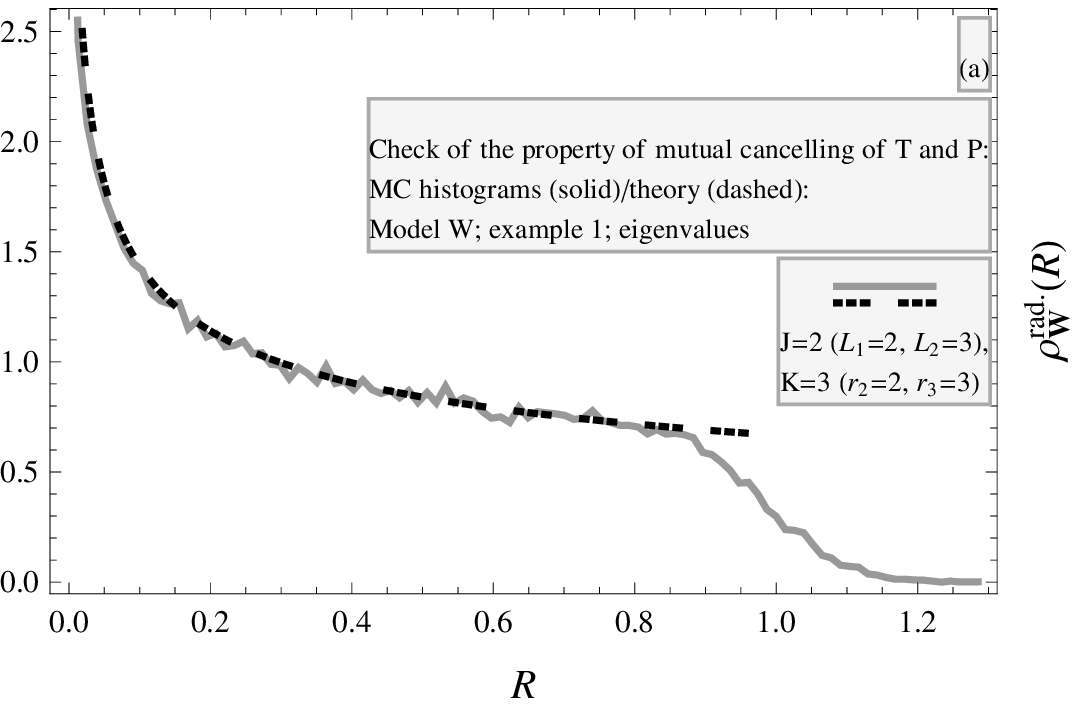}
\includegraphics[width=\columnwidth]{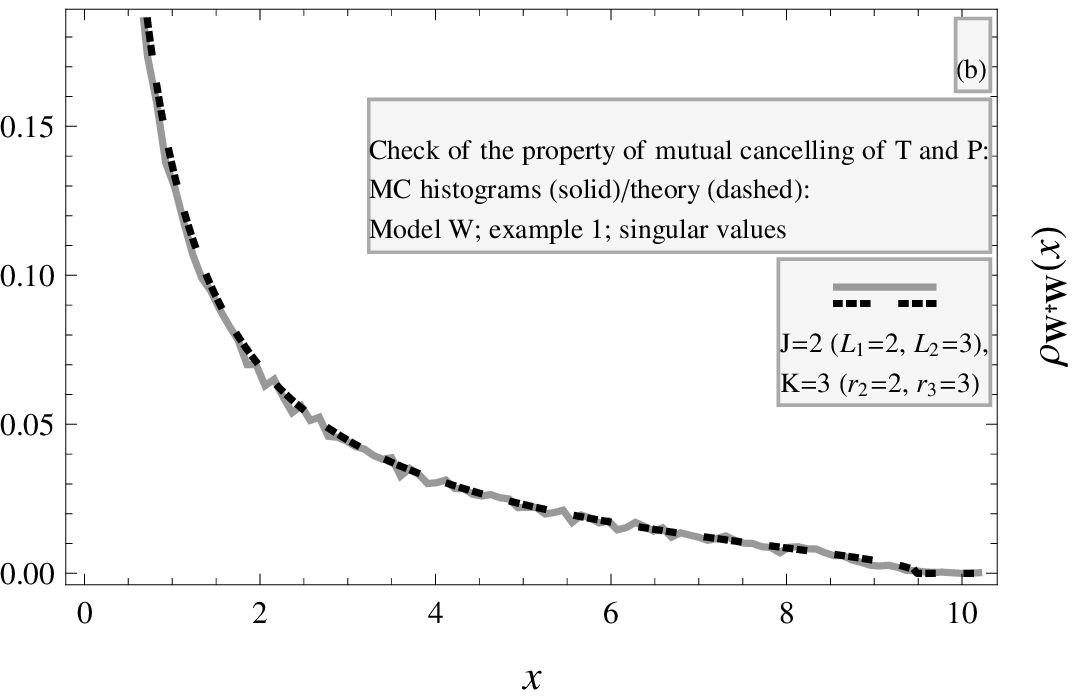}
\caption{Numerical verification of the property described in \emph{Remark 2} in Sec.~\ref{sss:WExample1}. The theoretical graphs are the level densities of the model \smash{$\mathbf{P}_{1}$}, obtained from (\ref{eq:HolomorphicNTransformOfPDaggerP}) with all \smash{$r_{k} = 1$}.}
\label{fig:ModelWInv}
\end{figure*}

\begin{figure*}[t]
\includegraphics[width=\columnwidth]{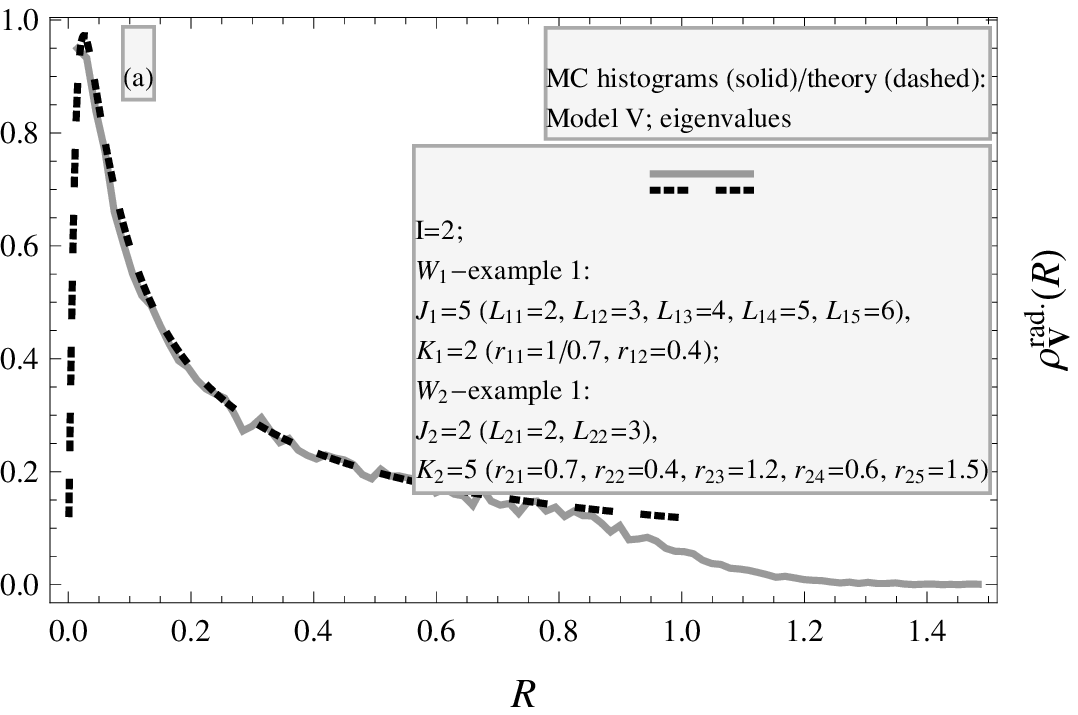}
\includegraphics[width=\columnwidth]{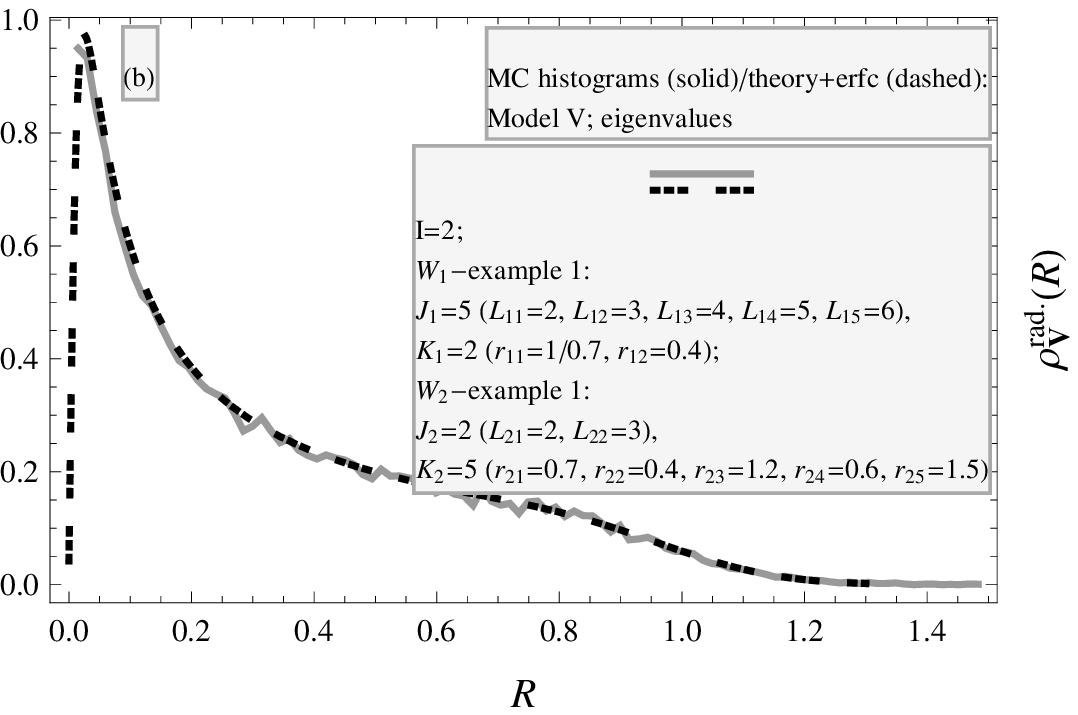}
\includegraphics[width=\columnwidth]{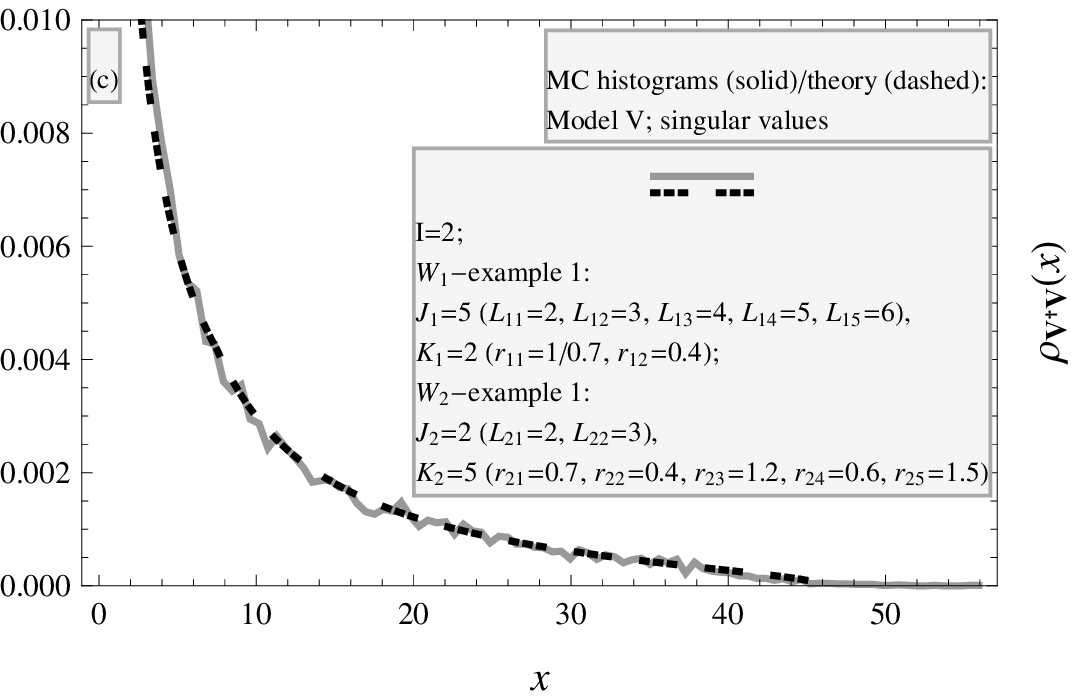}
\caption{Theoretical level densities [the eigenvalues (a), the eigenvalues plus the erfc form-factor (b), the singular values (c)] versus Monte Carlo data for the model $\mathbf{V}$.}
\label{fig:ModelV}
\end{figure*}

As the first example, take arbitrary $J$ and \smash{$L_{j}$}, but suppose that the weights for each \smash{$\mathbf{S}_{j}$} are equal to each other and denoted by
\begin{equation}\label{eq:TExample1Derivation01}
w_{j l} = \frac{w_{j}}{\sqrt{L_{j}}} , \quad l = 1 , 2 , \ldots , L_{j} ,
\end{equation}
for some $J$ constants \smash{$w_{j}$}.

\emph{Eigenvalues.} Then, the master equations (\ref{eq:TMasterEquation1})-(\ref{eq:TMasterEquation3}) give explicitly
\begin{equation}\label{eq:TExample1Derivation02}
\mathfrak{N}_{\mathbf{S}_{j}} ( z ) = \left| w_{j} \right|^{2} \frac{z + 1}{\frac{z}{L_{j}} + 1} , \quad j = 1 , 2 , \ldots , J ,
\end{equation}
which inserted into (\ref{eq:TMasterEquation0}) and after performing the functional inversion (\ref{eq:RotationallySymmetricNonHolomorphicNTransformDefinition}) leads to a polynomial equation of order $J$ for $\mathfrak{M}$,
\begin{equation}\label{eq:TExample1Derivation03}
\frac{R^{2}}{| w |^{2}} = \frac{( \mathfrak{M} + 1 )^{J}}{\prod_{j = 1}^{J} \left( \frac{\mathfrak{M}}{L_{j}} + 1 \right)} ,
\end{equation}
where for short, \smash{$w \equiv \prod_{j = 1}^{J} w_{j}$}. [Notice a certain similarity to the counterpart Eq.~(\ref{eq:HolomorphicNTransformOfPDaggerP}) for $\mathbf{P}$; indeed, I will show (Sec.~\ref{sss:WExample1}) that $\mathbf{T}$ of Example 1 is in some sense ``inverse'' to a certain class of models $\mathbf{P}$.]

Inserting $\mathfrak{M} = 0$ or $\mathfrak{M} = - 1$ (\ref{eq:SRExt})-(\ref{eq:SRInt}) (no zero modes here, $\alpha = 0$) into (\ref{eq:TExample1Derivation03}), one finds that the mean spectral domain is a centered disk of radius
\begin{equation}\label{eq:TExample1Derivation04}
R_{\textrm{ext.}} = | w | .
\end{equation}

In particular, if all \smash{$L_{j} = L$}, Eq.~(\ref{eq:TExample1Derivation03}) can be solved explicitly, yielding (\ref{eq:RadialMeanSpectralDensityFromRotationallySymmetricNonHolomorphicMTransform}),
\begin{equation}\label{eq:TExample1Derivation05}
\rho^{\textrm{rad.}}_{\mathbf{T}} ( R ) = 2 | w |^{2} \frac{1}{J} \left( 1 - \frac{1}{L} \right) \frac{R^{2 / J - 1}}{\left( | w |^{2} - \frac{R^{2 / J}}{L} \right)^{2}} ,
\end{equation}
for \smash{$R \leq | w |$}, and zero otherwise. This could also be obtained by using the scaling relation (\ref{eq:TEqualTermsScalingRelation}) and the proper result for $\mathbf{S}$, i.e., Eq.~(58) of~\cite{Jarosz2011-01}.

\emph{Singular values.} The master equation for the holomorphic $M$-transform \smash{$M \equiv M_{\mathbf{T}^{\dagger} \mathbf{T}} ( z )$} is a small modification of (\ref{eq:TExample1Derivation03}) according to (\ref{eq:NTransformConjecture}),
\begin{equation}\label{eq:TExample1Derivation06}
\frac{z}{| w |^{2}} = \frac{( M + 1 )^{J + 1}}{M \prod_{j = 1}^{J} \left( \frac{M}{L_{j}} + 1 \right)} ,
\end{equation}
which is polynomial of order $( J + 1 )$.

The above findings, along with the erfc conjecture (\ref{eq:FiniteSizeBorderlineFactor}), are tested against Monte Carlo data in Fig.~\ref{fig:ModelT} [(a), (b), (c)].

\emph{Remark 1: Divergence at zero.} Let me note that from the above equations one may easily derive an interesting feature of the level densities---their divergence close to zero. I will apply the logic presented in~\cite{BurdaJaroszLivanNowakSwiech20102011} to this and the following models. Assume that for $z \to 0$, also $z \mathfrak{G} \to 0$ and $z G \to 0$ (recall, $\mathfrak{M} = z \mathfrak{G} - 1$, $M = z G - 1$), which will be verified a posteriori. Then the denominators of the right hand sides of [(\ref{eq:TExample1Derivation03}), (\ref{eq:TExample1Derivation06})] tend to nonzero constants, and consequently, \smash{$\mathfrak{G} \sim z^{1 / J - 1} \overline{z}^{1 / J}$}, \smash{$G \sim z^{- J / ( J + 1 )}$}, i.e.,
\begin{subequations}
\begin{align}
\rho^{\textrm{rad.}}_{\mathbf{T}} ( R ) &\sim R^{- \frac{d - 2}{d}} , \quad R \to 0 ,\label{eq:TExample1Derivation07a}\\
\rho_{\mathbf{T}^{\dagger} \mathbf{T}} ( x ) &\sim x^{- \frac{d}{d + 1}} , \quad x \to 0 ,\label{eq:TExample1Derivation07b}
\end{align}
\end{subequations}
where
\begin{equation}\label{eq:TExample1Derivation08}
d = J .
\end{equation}
[The initial suppositions thus hold true: \smash{$z \mathfrak{G} \sim R^{2 / J} \to 0$} and \smash{$z G \sim z^{1 / ( J + 1 )} \to 0$}.] Note that the mean density of the singular values diverges for any $J$, while for the eigenvalues, the radial mean spectral density is finite for $J = 1$ (i.e., the model $\mathbf{S}$, for which it zeroes) or $J = 2$, in which case however, the proper density (\ref{eq:NonHermitianMeanSpectralDensityDefinition}) diverges.

\emph{Remark 2: Is $\mathbf{T}$ really different from $\mathbf{S}$?} Notice that if one opens up the brackets in the definition of the model $\mathbf{T}$ [(\ref{eq:TDefinition}), (\ref{eq:SDefinition})] (let me focus on this Example 1 and all \smash{$w_{j} = 1$}), \smash{$\mathbf{T} = \prod_{j = 1}^{J} ( L_{j}^{- 1 / 2} \sum_{l = 1}^{L_{j}} \mathbf{U}_{j l} )$}, one obtains a sum of \smash{$L \equiv \prod_{j = 1}^{J} L_{j}$} terms (multiplied by \smash{$L^{- 1 / 2}$}), each of which is a product of some $J$ CUE matrices. Now, it is known that a product of CUE matrices still belongs to the CUE, hence, $\mathbf{T}$ looks like the model $\mathbf{S}$ of length $L$---except the fact that now the various terms are not statistically independent (free). So one might ask how relevant these correlations between the $L$ terms are, i.e., how much the level densities of $\mathbf{T}$ and of $\mathbf{S}$ of length $L$ differ. Figure~\ref{fig:ModelTCorrelatedU} illustrates the theoretical mean spectral densities of these two random matrix models---they look completely different, showing that even though $\mathbf{T}$ may be recast as a sum of CUE matrices, correlations between them are important, and it is not our model $\mathbf{S}$ at all. Especially, the behavior at zero of the level densities in these two cases is entirely distinct, cf.~\emph{Remark 1} above. Also, with growing $L$, the density of $\mathbf{S}$ tends to that of the square GinUE distribution (cf.~\cite{Jarosz2011-01}), $\rho^{\textrm{rad.}}_{\mathbf{S}} ( R ) \to 2 R$ inside the unit circle and zero otherwise, while the limit of $\mathbf{T}$ is very different.


\subsubsection{Example 2}
\label{sss:TExample2}

As the second example, take arbitrary $J$, and let all the lengths \smash{$L_{j} = 2$}, but with arbitrary weights \smash{$w_{j 1}$}, \smash{$w_{j 2}$}.

\emph{Eigenvalues.} In this case, the master equations (\ref{eq:TMasterEquation1})-(\ref{eq:TMasterEquation3}) can be transformed into
\begin{equation}\label{eq:TExample2Derivation01}
z = \frac{1}{\frac{\left( \left| w_{j 1} \right| + \left| w_{j 2} \right| \right)^{2}}{\mathfrak{N}_{\mathbf{S}_{j}} ( z )} - 1} + \frac{1}{\frac{\left( \left| w_{j 1} \right| - \left| w_{j 2} \right| \right)^{2}}{\mathfrak{N}_{\mathbf{S}_{j}} ( z )} - 1} ,
\end{equation}
which are quadratic for \smash{$\mathfrak{N}_{\mathbf{S}_{j}} ( z )$}; they are supplied by the multiplication law (\ref{eq:TMasterEquation0}).

The borderline of the mean spectral domain is in general a centered annulus of radii (\ref{eq:SRExt})-(\ref{eq:SRInt}),
\begin{subequations}
\begin{align}
R_{\textrm{ext.}}^{2} &= \prod_{j = 1}^{J} \left( \left| w_{j 1} \right|^{2} + \left| w_{j 2} \right|^{2} \right) ,\label{eq:TExample2Derivation02a}\\
R_{\textrm{int.}}^{2} &= \prod_{j = 1}^{J} \left| \left| w_{j 1} \right|^{2} - \left| w_{j 2} \right|^{2} \right| ,\label{eq:TExample2Derivation02b}
\end{align}
\end{subequations}
which reduces to a disk only if the (absolute values of the) two weights in at least one term are equal to each other.

\emph{Singular values.} As above, with the left-hand side of (\ref{eq:TMasterEquation0}) changed according to (\ref{eq:NTransformConjecture}).

These master equations are solved numerically and compared to the Monte Carlo eigenvalues in Fig.~\ref{fig:ModelT} [(d), (e), (f)].

\emph{Remark: Divergence at zero.} Assume that there is a certain number $1 \leq \mathcal{W} \leq J$ of terms \smash{$\mathbf{S}_{j}$} whose (absolute values of the) two weights are equal to each other, \smash{$| w_{j 1} | = | w_{j 2} |$}; then one may ask about a behavior of the above level densities close to zero. For each of these $\mathcal{W}$ terms, Eq.~(\ref{eq:TExample2Derivation01}) becomes linear and yields (I already replace $z$ by $\mathfrak{M} = z \mathfrak{G} - 1$ or $M = z G - 1$, respectively, and assume $z \mathfrak{G} \to 0$ and $z G \to 0$ for $z \to 0$), \smash{$\mathfrak{N}_{\mathbf{S}_{j}} ( \mathfrak{M} ) = 4 | w_{j} |^{2} z \mathfrak{G} / ( z \mathfrak{G} + 1 )$}, which tends to zero as $z \mathfrak{G}$ for $z \to 0$. For the remaining $( J - \mathcal{W} )$ terms in $\mathbf{T}$, Eq.~(\ref{eq:TExample2Derivation01}) implies that any \smash{$\mathfrak{N}_{\mathbf{S}_{j}} ( \mathfrak{M} )$} tends to a nonzero constant \smash{$| | w_{j 1} |^{2} - | w_{j 2} |^{2} |$} for $z \to 0$. Substituting these results into (\ref{eq:TMasterEquation0}), one discovers that the densities diverge at zero again according to (\ref{eq:TExample1Derivation07a})-(\ref{eq:TExample1Derivation07b}), but with
\begin{equation}\label{eq:TExample2Derivation03}
d = \mathcal{W} .
\end{equation}
(This finding is consistent with $z \mathfrak{G} \to 0$ and $z G \to 0$ for $z \to 0$.)


\subsection{Product $\mathbf{W}$}
\label{ss:ProductW}


\subsubsection{Master equations for $\mathbf{W}$}
\label{sss:MasterEquationsForW}

\emph{Eigenvalues.} The master equations for the mean spectral density of $\mathbf{W}$ (\ref{eq:WDefinition}) are easily obtained through the multiplication law (\ref{eq:EigenvaluesOfTheProductXAssumingAllXiAreSquareAndHaveRotationallySymmetricMeanSpectraDerivation01}), \smash{$\mathfrak{N}_{\mathbf{W}} ( z ) = \mathfrak{N}_{\mathbf{T}} ( z ) \mathfrak{N}_{\mathbf{P}} ( z )$}, from formula (\ref{eq:HolomorphicNTransformOfPDaggerP}) (obviously with \smash{$r_{1} = 1$})---they read
\begin{equation}\label{eq:WMasterEquation0}
\mathfrak{N}_{\mathbf{W}} ( z ) = \sigma^{2} ( z + 1 ) \prod_{k = 2}^{K} \left( \frac{z}{r_{k}} + 1 \right) \mathfrak{N}_{\mathbf{T}} ( z ) ,
\end{equation}
where the rotationally-symmetric nonholomorphic $N$-transform of $\mathbf{T}$ is given by (\ref{eq:TMasterEquation0})-(\ref{eq:TMasterEquation3}); one should substitute here \smash{$\mathfrak{M} \equiv \mathfrak{M}_{\mathbf{W}} ( R^{2} )$} in the place of $z$.

\emph{Singular values.} Choose arbitrary \smash{$r_{1}$}, use the multiplication law (\ref{eq:SingularValuesOfTheProductXViaCyclicShiftsAndMultiplicationLawDerivation05}) along with (\ref{eq:HolomorphicNTransformOfPDaggerP})---obtaining,
\begin{equation}\label{eq:WdWMasterEquation0}
N_{\mathbf{W}^{\dagger} \mathbf{W}} ( z ) = \frac{\sigma^{2}}{\sqrt{r_{1}}} ( z + 1 ) \prod_{k = 2}^{K} \left( \frac{z}{r_{k}} + 1 \right) N_{\mathbf{T}^{\dagger} \mathbf{T}} \left( \frac{z}{r_{1}} \right) ,
\end{equation}
where the holomorphic $N$-transform of \smash{$\mathbf{T}^{\dagger} \mathbf{T}$} is given by [(\ref{eq:NTransformConjecture}), (\ref{eq:TMasterEquation0})-(\ref{eq:TMasterEquation3})]; one should replace here $z$ by \smash{$M \equiv M_{\mathbf{W}^{\dagger} \mathbf{W}} ( z )$}.


\subsubsection{Example 1}
\label{sss:WExample1}

Consider the situation described in Sec.~\ref{sss:TExample1} [i.e., arbitrary $J$ and \smash{$L_{j}$}, weights for each \smash{$\mathbf{S}_{j}$} equal to each other (\ref{eq:TExample1Derivation01})] plus arbitrary rectangularity ratios \smash{$r_{k}$}.

\emph{Eigenvalues.} The master equation is polynomial of order $( J + K )$,
\begin{equation}\label{eq:WExample1Derivation01}
\frac{R^{2}}{| w |^{2} \sigma^{2}} = ( \mathfrak{M} + 1 )^{J + 1} \frac{\prod_{k = 2}^{K} \left( \frac{\mathfrak{M}}{r_{k}} + 1 \right)}{\prod_{j = 1}^{J} \left( \frac{\mathfrak{M}}{L_{j}} + 1 \right)} ,
\end{equation}
valid inside the disk of radius (\ref{eq:XRExt})-(\ref{eq:XRInt}),
\begin{equation}\label{eq:WExample1Derivation02}
R_{\textrm{ext.}} = | w | \sigma .
\end{equation}
[The internal radius vanishes because the density of the zero modes, \smash{$\alpha = 1 - \min \{ r_{k} \}$}, hence, either $\mathfrak{M} = - 1$ or at least one term in the product over $k$ in (\ref{eq:WExample1Derivation01}) with $\mathfrak{M} = \alpha - 1$ is zero. This same argument will be valid in all the models investigated henceforth.]

\emph{Singular values.} The master equation is polynomial of order $( J + K + 1 )$,
\begin{equation}\label{eq:WExample1Derivation03}
\frac{z}{| w |^{2} \sigma^{2}} = \sqrt{r_{1}} \frac{( M + 1 ) \left( \frac{M}{r_{1}} + 1 \right)^{J + 1}}{M} \frac{\prod_{k = 2}^{K} \left( \frac{M}{r_{k}} + 1 \right)}{\prod_{j = 1}^{J} \left( \frac{M}{r_{1} L_{j}} + 1 \right)} .
\end{equation}

Figures~\ref{fig:ModelWK2} [(a), (b), (c)] and~\ref{fig:ModelWK5} [(a), (b), (c)] present numerical solutions to these equations and the corresponding Monte Carlo results, both in perfect agreement.

\emph{Remark 1: Divergence at zero.} Following the same line of reasoning as in \emph{Remark 1} in Sec.~\ref{sss:TExample1}, one recognizes that the divergences close to zero of the level densities stemming from [(\ref{eq:WExample1Derivation01}), (\ref{eq:WExample1Derivation03})]---disregarding their possible zero-mode part, which amounts to considering \smash{$r_{1} \geq 1$}, which means that there cannot happen \smash{$L_{j} = 1 / r_{1}$}, i.e., one may disregard in our analysis the products over $j$ in the above formulae---are governed by the number \smash{$\mathcal{R} \equiv \# \left\{ 2 \leq k \leq K : r_{k} = 1 \right\}$}, and given again by (\ref{eq:TExample1Derivation07a})-(\ref{eq:TExample1Derivation07b}) but with
\begin{equation}\label{eq:WExample1Derivation04}
d = ( J + 1 ) \delta_{r_{1} , 1} + \mathcal{R} .
\end{equation}

\emph{Remark 2: $\mathbf{T}$ as an ``inverse'' of $\mathbf{P}$ with integer rectangularity ratios.} The above master equations of $\mathbf{W}$ imply the following peculiar joint property of the models $\mathbf{T}$ (of Example 1) and $\mathbf{P}$: Assume that $\mathbf{P}$ (of length $K$; set for simplicity $\sigma = 1$) is such that a certain number $1 \leq J \leq K$ of the rectangularity ratios \smash{$r_{k} / r_{1}$} are integers greater than $1$. If one multiplies this $\mathbf{P}$ from the left by a product $\mathbf{T}$ of length $J$ such that the lengths \smash{$L_{j}$} of its terms are equal to these integer rectangularity ratios (and \smash{$w_{j} = 1$}), then the master equations [(\ref{eq:WExample1Derivation01}), (\ref{eq:WExample1Derivation03})] turn into
\begin{equation}\label{eq:WExample1Derivation05}
R^{2} = ( \mathfrak{M} + 1 )^{J + 1} \prod_{\substack{2 \leq k \leq K :\\\textrm{$r_{k}$ is not integer $> 1$}}} \left( \frac{\mathfrak{M}}{r_{k}} + 1 \right) ,
\end{equation}
\begin{equation}
\begin{split}\label{eq:WExample1Derivation06}
z &= \sqrt{r_{1}} \frac{( M + 1 ) \left( \frac{M}{r_{1}} + 1 \right)^{J + 1}}{M} \cdot\\
&\cdot \prod_{\substack{2 \leq k \leq K :\\\textrm{$r_{k} / r_{1}$ is not integer $> 1$}}} \left( \frac{M}{r_{k}} + 1 \right) .
\end{split}
\end{equation}
Comparing [(\ref{eq:WExample1Derivation05}), (\ref{eq:WExample1Derivation06})] with (\ref{eq:HolomorphicNTransformOfPDaggerP}), one recognizes that they are the master equations for a model \smash{$\widetilde{\mathbf{P}} \equiv \mathbf{P}_{1} \mathbf{P}_{2}$}, where \smash{$\mathbf{P}_{1} \equiv \mathbf{B}_{1} \ldots \mathbf{B}_{J}$}, with \smash{$\mathbf{B}_{j}$} being the square GinUE random matrices (\ref{eq:RectangularGinUEJPDF}), while \smash{$\mathbf{P}_{2}$} is the model $\mathbf{P}$ with the terms \smash{$\mathbf{A}_{k}$} for which \smash{$r_{k} / r_{1} = L_{j}$} removed from it.

In particular, if the model $\mathbf{P}$ is such that all of its rectangularity ratios \smash{$r_{k} / r_{1}$}, $k \geq 2$, are integers greater than $1$, then multiplying it from the left by the appropriate matrix $\mathbf{T}$ [of length $( K - 1 )$ and the lengths of its terms \smash{$L_{k} = r_{k} / r_{1}$}] yields
\begin{equation}\label{eq:WExample1Derivation07}
\mathfrak{M} = R^{2 / K} - 1 , \quad \textrm{i.e.,} \quad \rho^{\textrm{rad.}}_{\mathbf{W}} ( R ) = \frac{2}{K} R^{2 / K - 1} ,
\end{equation}
\begin{equation}\label{eq:WExample1Derivation08}
z = \sqrt{r_{1}} \frac{( M + 1 ) \left( \frac{M}{r_{1}} + 1 \right)^{K}}{M} .
\end{equation}

In other words, the model $\mathbf{T}$ (of Example 1) is ``inverse'' (at least on the level of the mean densities of eigenvalues and singular values) to the model $\mathbf{P}$ with rectangularity ratios \smash{$r_{k} / r_{1}$}, $k \geq 2$, equal to the lengths \smash{$L_{j}$}, with the ``unity'' being \smash{$\mathbf{P}_{1}$} [(\ref{eq:WExample1Derivation07}), (\ref{eq:WExample1Derivation08})]. These results are verified numerically in Figs.~\ref{fig:ModelWInv} [(a), (b)].


\subsubsection{Example 2}
\label{sss:WExample2}

In the setting of Sec.~\ref{sss:TExample2} [arbitrary $J$, all \smash{$L_{j} = 2$}, arbitrary weights \smash{$w_{j 1}$}, \smash{$w_{j 2}$}] with arbitrary \smash{$r_{k}$}, the master equations (\ref{eq:WMasterEquation0}) and (\ref{eq:WdWMasterEquation0}) should be supplied by the result (\ref{eq:TExample2Derivation01}). These can be solved numerically and are compared to the numerical simulations in Figs.~\ref{fig:ModelWK2} [(d), (e), (f)] and~\ref{fig:ModelWK5} [(d), (e), (f)].

An analytical result can be obtained for the mean spectral domain---it is a disk of radius (\ref{eq:XRExt})-(\ref{eq:XRInt}),
\begin{equation}\label{eq:WExample2Derivation01}
R_{\textrm{ext.}}^{2} = \sigma^{2} \prod_{j = 1}^{J} \left( \left| w_{j 1} \right|^{2} + \left| w_{j 2} \right|^{2} \right) .
\end{equation}
This should be contrasted with the analogous situation for the model $\mathbf{T}$, where the domain is in general an annulus (\ref{eq:TExample2Derivation02a})-(\ref{eq:TExample2Derivation02b}).

\emph{Remark: Divergence at zero.} Combining the reasonings from Secs.~\ref{sss:TExample2} and~\ref{sss:WExample1}, one finds that for the mean spectral domains to touch zero, there must be a number $\mathcal{W}$ of weights obeying \smash{$| w_{j 1} | = | w_{j 2} |$}, in which case the divergences at zero of the level densities (without taking into account possible zero modes) are still described by (\ref{eq:TExample1Derivation07a})-(\ref{eq:TExample1Derivation07b}) but with
\begin{equation}\label{eq:WExample2Derivation02}
d = ( \mathcal{W} + 1 ) \delta_{r_{1} , 1} + \mathcal{R} .
\end{equation}
Notice that this is consistent with (\ref{eq:WExample1Derivation05}), and even suggests how an expression for $d$ might look like for a most arbitrary model $\mathbf{W}$.


\subsection{Product $\mathbf{V}$}
\label{ss:ProductV}


\subsubsection{Master equations for $\mathbf{V}$}
\label{sss:MasterEquationsForV}

The mean densities of the eigenvalues or singular values of any product $\mathbf{V}$ (\ref{eq:VDefinition}) can be directly derived from (\ref{eq:WMasterEquation0}) or (\ref{eq:WdWMasterEquation0}) using the multiplication laws (\ref{eq:EigenvaluesOfTheProductXAssumingItIsSquareAndHasRotationallySymmetricMeanSpectrumDerivation01}) [or (\ref{eq:EigenvaluesOfTheProductXAssumingAllXiAreSquareAndHaveRotationallySymmetricMeanSpectraDerivation01})] or (\ref{eq:SingularValuesOfTheProductXViaCyclicShiftsAndMultiplicationLawDerivation06}).


\subsubsection{Example}
\label{sss:VExample1}

I will consider in this paper just one instance of the model $\mathbf{V}$, namely, any length $I$ but all the terms \smash{$\mathbf{W}_{i} = \mathbf{T}_{i} \mathbf{P}_{i}$} of the form of Example 1 above (Sec.~\ref{sss:WExample1}) [i.e., the lengths \smash{$J_{i}$} of \smash{$\mathbf{T}_{i}$} arbitrary, with arbitrary lengths \smash{$L_{i j}$} of \smash{$\mathbf{S}_{i j}$} but with the weights \smash{$w_{i j l} = w_{i j} / \sqrt{L_{i j}}$} independent of $l$; the lengths \smash{$K_{i}$} of \smash{$\mathbf{P}_{i}$}, variances \smash{$\sigma_{i k}^{2}$} and rectangularity ratios \smash{$r_{i k}$} of \smash{$\mathbf{A}_{i k}$} also arbitrary], having moreover arbitrary rectangularity ratios (\ref{eq:sDefinition}), \smash{$s_{i} = N_{i 1} / N_{I , K_{I} + 1} = \prod_{i^{\prime} = i}^{I} r_{i^{\prime} 1}$} (of course, \smash{$s_{1} = 1$} if one investigates the eigenvalues of $\mathbf{V}$).

\emph{Eigenvalues.} The master equation is polynomial of order \smash{$\sum_{i = 1}^{I} ( J_{i} + K_{i} )$},
\begin{equation}\label{eq:VExample1Derivation01}
\frac{R^{2}}{| w |^{2} \sigma^{2}} = \prod_{i = 1}^{I} \left( \frac{\mathfrak{M}}{s_{i}} + 1 \right)^{J_{i} + 1} \frac{\prod_{k = 2}^{K_{i}} \left( \frac{\mathfrak{M}}{s_{i + 1} r_{i k}} + 1 \right)}{\prod_{j = 1}^{J_{i}} \left( \frac{\mathfrak{M}}{s_{i} L_{i j}} + 1 \right)} ,
\end{equation}
where \smash{$\sigma \equiv \prod_{i = 1}^{I} \prod_{k = 2}^{K_{i}} \sigma_{i k}$} and \smash{$w \equiv \prod_{i = 1}^{I} \prod_{j = 1}^{J_{i}} w_{i j}$}, and it is valid inside the disk of radius (\ref{eq:XRExt})-(\ref{eq:XRInt}),
\begin{equation}\label{eq:VExample1Derivation02}
R_{\textrm{ext.}} = | w | \sigma .
\end{equation}

\emph{Singular values.} The master equation is polynomial of order \smash{$\sum_{i = 1}^{I} ( J_{i} + K_{i} ) + 1$},
\begin{equation}
\begin{split}\label{eq:VExample1Derivation03}
\frac{z}{| w |^{2} \sigma^{2}} &= \sqrt{s_{1}} \frac{M + 1}{M} \cdot\\
&\cdot \prod_{i = 1}^{I} \left( \frac{M}{s_{i}} + 1 \right)^{J_{i} + 1} \frac{\prod_{k = 2}^{K_{i}} \left( \frac{M}{s_{i + 1} r_{i k}} + 1 \right)}{\prod_{j = 1}^{J_{i}} \left( \frac{M}{s_{i} L_{i j}} + 1 \right)} .
\end{split}
\end{equation}

Figures~\ref{fig:ModelV} [(a), (b), (c)] illustrate numerical solutions to these polynomial equations and test them against Monte Carlo simulations, finding perfect agreement.

\emph{Remark 1: Generalized Bures products.} Notice that for $I = 1$, Eqs.~[(\ref{eq:VExample1Derivation01}), (\ref{eq:VExample1Derivation03})] reduce to [(\ref{eq:WExample1Derivation01}), (\ref{eq:WExample1Derivation03})]. Also, setting $I = 1$, \smash{$J_{1} = 1$}, \smash{$K_{1} = 1$}, $w = 1$, $\sigma = 1$, \smash{$s_{1} = r_{1} = 1$}, \smash{$L_{1 1} = 2$} in (\ref{eq:VExample1Derivation03}) yields a cubic equation, whose solution is the original Bures distribution (\ref{eq:BuresDistribution}). Of course, an even broader generalization of the Bures model would be to include distinct weights \smash{$w_{i j l}$} in the \smash{$\mathbf{S}_{i j}$}; I refrain from explicitly writing down the pertinent master equations, even though it is a straightforward step from the above results.

\emph{Remark 2: Divergence at zero.} The level densities stemming from [(\ref{eq:VExample1Derivation01}), (\ref{eq:VExample1Derivation03})] diverge at zero (beyond possible zero modes) again according to (\ref{eq:TExample1Derivation07a})-(\ref{eq:TExample1Derivation07b}) but with
\begin{equation}\label{eq:VExample1Derivation04}
d = \sum_{i = 1}^{I} \left( \left( J_{i} + 1 \right) \delta_{s_{i} , 1} + \mathcal{R}_{i} \right)  ,
\end{equation}
where for short, \smash{$\mathcal{R}_{i} \equiv \# \left\{ 2 \leq k \leq K_{i} : r_{i k} = 1 / s_{i + 1} \right\}$}. Formula (\ref{eq:VExample1Derivation04}) reduces to (\ref{eq:WExample1Derivation04}) for $I = 1$, as it should.


\section{Conclusions}
\label{s:Conclusions}


\subsection{Summary}
\label{ss:Summary}

In this paper, I attempted to show how free probability theory (the multiplication law in the realm of Hermitian random matrices and the addition law for non-Hermitian ones) greatly simplifies otherwise difficult calculations of the mean densities of eigenvalues and singular values of the generalized Bures products $\mathbf{T}$, $\mathbf{W}$, $\mathbf{V}$, which are random matrix models of relevance in quantum information theory.


\subsection{Open problems}
\label{ss:OpenProblems}

As already mentioned, this article is certainly only the initial step (nevertheless important, especially from the point of view of quantum information theory) in learning about the models $\mathbf{T}$, $\mathbf{W}$, $\mathbf{V}$. A major endeavor would be to consider finite matrix dimensions and attempt a computation of the complete JPDF of the eigenvalues of these models. One could also investigate some of their universal properties (cf.~Sec.~\ref{sss:MeanDensitiesOfTheEigenvaluesAndSingularValues}). Actually, inspired by~\cite{BurdaJanikWaclaw2009}, one could check whether the level densities themselves show any sign of universality, just like for $\mathbf{P}$.

A pressing challenge is to prove the three hypotheses which my derivation is founded upon: the $N$-transform conjecture (cf.~Sec.~\ref{sss:RotationalSymmetryAndNTransformConjecture}), the single ring conjecture (cf.~Sec.~\ref{sss:SingleRingConjecture}), and the erfc conjecture (cf.~Sec.~\ref{sss:ErfcConjecture}).

It would also be desirable to understand more about the features of our models important for quantum entanglement theory, e.g. analyze their von Neumann entropy (\ref{eq:VonNeumannEntropy}).


\begin{acknowledgments}
My work has been partially supported by the Polish Ministry of Science and Higher Education Grant ``Iuventus Plus'' No.~0148/H03/2010/70. I acknowledge the financial support of Clico Ltd., Oleandry 2, 30-063 Krak\'{o}w, Poland, while completing parts of this paper. I am grateful to Karol \.{Z}yczkowski for valuable discussions.
\end{acknowledgments}



\end{document}